\documentclass[12pt]{article}
\usepackage{amssymb,amsmath}
\makeatletter

\usepackage{arydshln}

\newcount\comment@nesting

\begingroup
\xdef\comment@begincomment{\string\\begin\string\{comment\string\}}
\xdef\comment@endcomment{\string\\end\string\{comment\string\}}
\endgroup

\begingroup
\catcode`\^^M=\active
\def\@temp{\endgroup\def\comment@processline##1^^M}%
\@temp{%
    \def\comment@curline{#1}%
    \let\@next=\comment@processline
    \ifx\comment@curline\comment@endcomment
        \ifnum\comment@nesting=0
            \def\@next{\end{comment}}%
        \else
            \global\advance\comment@nesting by -1
        \fi
    \else
        \ifx\comment@curline\comment@begincomment
            \global\advance\comment@nesting by 1
        \fi
    \fi
    \@next
}

\makeatother

\usepackage[dvipdfmx]{graphicx}
\usepackage{bm}
\usepackage{color}
\usepackage{here}
\usepackage{enumerate}
\usepackage{here}
\usepackage{subfigure}
\usepackage{tikz}
\usepackage{hyperref}
\newcommand{\Zb}{\mathbb{Z}}

\newcommand{\Acal}{\mathcal{A}}

\newcommand{\Ncal}{\mathcal{N}}

\DeclareMathOperator*{\Tr}{{\rm Tr}}

\renewcommand{\Im}{\operatorname{Im}}

\numberwithin{equation}{section}

\allowdisplaybreaks[1]

\definecolor{mygreen}{rgb}{0,0.714,0.286}

\begin{document}

%%%%%%%%%%%%%%%%%%%%%%%%%%%%%%%%%%%%%%%%%%%%
\thispagestyle{empty}
\begin{flushright}
%%%%%%%%%%%%%%%%%%%%%%%%%%%%%%%%%%%%%%%%%%%%%%%%%%%%%%%%%%%%%%%%%
%input \\
%%%%%%%%%%%%%%%%%%%%%%%%%%%%%%%%%%%%%%%%%%%%%%%%%%%%%%%%%%%%%%%%%

\end{flushright}
\vskip1.5cm
\begin{center}
{\Large \bf 
Line defect half-indices of 
\\
\vskip0.75cm
$SU(N)$ Chern-Simons theories
}

\vskip1.5cm
Tadashi Okazaki\footnote{tokazaki@seu.edu.cn}

\bigskip
{\it Shing-Tung Yau Center of Southeast University,\\
Yifu Architecture Building, No.2 Sipailou, Xuanwu district, \\
Nanjing, Jiangsu, 210096, China
}

\bigskip
and
\\
\bigskip
Douglas J. Smith\footnote{douglas.smith@durham.ac.uk}

\bigskip
{\it Department of Mathematical Sciences, Durham University,\\
Upper Mountjoy, Stockton Road, Durham DH1 3LE, UK}

\end{center}

%%%%%%%%%%%%%%%%%%%%%%%%%%%%%%%%%%%%%%%%%%%%
\vskip1cm
\begin{abstract}
We study the Wilson line defect half-indices of 3d $\mathcal{N}=2$ supersymmetric $SU(N)$ Chern-Simons theories of level $k\le -N$ 
with Neumann boundary conditions for the gauge fields, together with 2d Fermi multiplets and fundamental 3d chiral multiplets to cancel the gauge anomaly. We derive some exact results and also make some conjectures based on expansions of the $q$-series. We find several interesting connections with special functions known in the literature, including Rogers-Ramanujan functions for which we conjecture integral representations, and the appearance of Appell-Lerch sums for certain Wilson line half-index grand canonical ensembles which reveal an unexpected appearance of mock modular functions. We also find intriguing $q$-difference equations relating half-indices to Wilson line half-indices. Some of these results also have a description in terms of a dual theory with Dirichlet boundary conditions for the vector multiplet in the dual theory.
\end{abstract}
%%%%%%%%%%%%%%%%%%%%%%%%%%%%%%%%%%%%%%%%%%
\newpage
\setcounter{tocdepth}{3}
\tableofcontents
%%%%%%%%%%%%%%%%%%%%%%%%%%

%%%%%%%%%%%%%%%%%%%%%%%%%%%%%%%%%%
%%%%%%%%%%%%%%%%%%%%%%%%%%%%%%%%%%
\section{Introduction and conclusion}
\label{sec_intro}
%%%%%%%%%%%%%%%%%%%%%%%%%%%%%%%%%%
%%%%%%%%%%%%%%%%%%%%%%%%%%%%%%%%%%
In the presence of a boundary 3d $\mathcal{N}=2$ supersymmetric field theories 
can preserve $\mathcal{N}=(0,2)$ supersymmetry in such a way 
that the bulk fields satisfy certain boundary conditions. 
\footnote{See \cite{Gadde:2013wq,Okazaki:2013kaa,Gadde:2013sca,Yoshida:2014ssa,Dimofte:2017tpi,
Brunner:2019qyf,Costello:2020ndc,Sugiyama:2020uqh,Jockers:2021omw,Dedushenko:2021mds,Zeng:2021zef,
Okazaki:2021pnc,Okazaki:2021gkk,Okazaki:2023hiv,Okazaki:2023kpq} 
for various studies of $\mathcal{N}=(0,2)$ boundary conditions.} 
In addition, they can naturally couple to boundary 2d degrees of freedom preserving $\mathcal{N}=(0,2)$ supersymmetry. 
Consequently, such BPS boundary conditions can enlarge the web of non-trivial fixed points of 2d $\mathcal{N}=(0,2)$ SCFTs 
and the elliptic genera of 2d $\mathcal{N}=(0,2)$ supersymmetric theories \cite{Benini:2013nda,Benini:2013xpa,Gadde:2013dda}
are generalized to the half-indices of the 3d-2d coupled systems \cite{Gadde:2013wq, Gadde:2013sca, Yoshida:2014ssa,Dimofte:2017tpi}. 
The half-indices can be decorated by introducing the half-BPS line defect operators 
supported on a line perpendicular to the boundary preserving 1d $\mathcal{N}=2$ supersymmetry \cite{Dimofte:2017tpi}. 
When one expands such line defect half-indices of 1d-2d-3d system with respect to the fugacities, one can count BPS local operators living at the junctions of line defects and the boundary. 
Identities in the algebra of line defects lead to certain difference equations satisfied by half-indices as well as supersymmetric partition functions on a compact $3$-manifold \cite{Dimofte:2011ju,Beem:2012mb,Dimofte:2011py}. 

In this paper we study the line defect half-indices or equivalently correlation function of the line operators of 3d $\mathcal{N}=2$ supersymmetric $SU(N)_k$ Chern-Simons theories 
with level $k\le -N$ where the $SU(N)$ vector multiplet obeys Neumann boundary condition. 
As opposed to the cases with Dirichlet boundary condition for the $SU(N)$ vector multiplet, 
the (line defect) half-indices for such boundary conditions involve multi-dimensional matrix integrals and they are not studied very well so far. 
\footnote{See \cite{Dimofte:2017tpi} for the (line defect) half-indices for the Abelian cases 
and also for those with Dirichlet boundary condition for the $SU(N)$ vector multiplet, }

The quantum consistency requires the non-Abelian gauge anomaly to vanish. 
For $k = -N$, there is no gauge anomaly. 
While the half-index is trivial, 
the line defect half-index is non-trivial as it implements a detector of the BPS local operator attached with the end point of the Wilson lines, i.e.\ it counts the BPS operators in the conjugate representation of the gauge group to the Wilson line. We find exact results for the case of Wilson lines in symmetric rank-$k$ or charge-$n$ representations. We can also calculate a grand canonical ensemble of the charge-$n$ Wilson line half-indices and find that it has a simple expression in terms of Jacobi theta series or as a difference of Appell-Lerch sums, indicating an interesting mock modular property.

For $k$ being an integer with $k < -N$, 
the gauge anomaly cancellation can be naturally achieved by introducing $|k| - N$ boundary 2d $\mathcal{N}=(0,2)$ Fermi multiplets in the fundamental representation\footnote{Other options include replacing a 2d Fermi multiplet with a pair of 3d chiral multiplets in the fundamental or anti-fundamental representation with Dirichlet boundary conditions, or adding addition Fermi multiplets each with a pair of chirals with Neumann boundary conditions.}. 
The 3d bulk and 2d boundary system has a dual description as the Dirichlet boundary conditions of $U(|k|-N)_{|k|,N}$ pure Chern-Simons theory according to the level-rank duality \cite{Dimofte:2017tpi}.
We find the closed-form $q$-series expressions of the half-indices 
which can also be identified with the vacuum characters of the $U(|k|-N)_N$ WZW models \cite{Dimofte:2017tpi}.

Besides, we can also obtain the bulk-boundary system with $k$ a half-integer with $k < -N$ 
by coupling an odd number of 3d chiral multiplets in the fundamental or anti-fundamental representations obeying either Neumann or Dirichlet boundary conditions along with the required number of 2d Fermis to cancel the gauge anomaly\footnote{As with the case of integer $k$, a pair of such Dirichlet chirals can be replaced by a 2d Fermi multiplet and vice-versa, or additional Fermis can be added together with a pair of Neumann chirals.}. We mostly consider the case with a single fundamental chiral with Neumann boundary conditions and $|k| - N + \frac{1}{2}$ Fermi multiplets, which has a dual $U(|k| - N + \frac{1}{2})_{|k|, N}$ theory with a single fundamental chiral where both the vector multiplet and the chiral have Dirichlet boundary conditions.
This is closely related to the boundary dualities discussed in \cite{Dimofte:2017tpi}.

For $N=2$ and $3$ we analytically derive the exact closed-form expressions of half-indices using the Jacobi triple product identity. 
In particular, we find exact results for the half-indices and in the presence of symmetric rank-$k$ or charge-$n$ Wilson lines for $SU(2)_{-2}$, unflavored (the flavor fugacity $x \to 1$) $SU(2)_{-5/2}$ and $SU(2)_{-3}$. For $SU(3)_{-3}$ we derive the result with a charge-$n$ Wilson line and implicitly also the symmetric rank-$k$ case but we didn't present a derivation of this more lengthy calculation.
%N=2, k=-5/2
Remarkably, we find that for $SU(2)_{-5/2}$, the half-index and the one-point function of the Wilson line in the fundamental representation 
are identified with the Rogers-Ramanujan functions \cite{MR1117903}. 

%%%%%%%%%%%%%%%%%%%%%%%%%%%%%%%%%%
\subsection{Structure}
\label{sec_structure}
%%%%%%%%%%%%%%%%%%%%%%%%%%%%%%%%%%
The paper is organized as follows. 
In section \ref{sec_CS_suN} we discuss $\mathcal{N}=(0,2)$ supersymmetric boundary conditions 
for 3d $\mathcal{N}=2$ $SU(N)$ Chern-Simons-matter theories with fundamental chiral multiplets 
whose vector multiplet satisfies Neumann boundary condition. 
We briefly review line defect half-indices or equivalently correlation functions of line operators which decorates the half-indices. 
In section \ref{sec_su2} we study the half-indices and correlators for the $SU(2)$ Chern-Simons theories. 
Using the Jacobi triple product identity, we analytically obtain the closed-form expressions. 
In section \ref{sec_su3} we examine those for the $SU(3)$ Chern-Simons theories. 
In section \ref{sec_suN} we present conjectural formulas for general $SU(N)$ Chern-Simons theories. 

%%%%%%%%%%%%%%%%%%%%%%%%%%%%%%%%%%
\subsection{Future works}
\label{sec_future}
%%%%%%%%%%%%%%%%%%%%%%%%%%%%%%%%%%
There are several open problems 
which we hope to report in future works. 

\begin{itemize}
    \item While we focus on computing the line defect half-indices for gauge theories with $SU(N)$ gauge group and Chern-Simons level $k\le -N$ in this work, 
    it would be interesting to study the cases with different gauge groups, matter fields and Chern-Simons levels to figure out the $q$-difference equations. In particular, those with an adjoint chiral multiplet generalizes the dualities of $\mathcal{N}=(2,2)$ and $\mathcal{N}=(0,4)$ boundary conditions in \cite{Okazaki:2019bok,Okazaki:2020lfy}. 

    \item In our previous works \cite{Okazaki:2023hiv,Okazaki:2023kpq} we found the confining dualities of boundary conditions where the half-indices are identified with the Askey-Wilson $q$-beta integrals \cite{MR783216,MR772878,MR845667,MR1139492,MR1266569,MR2267266}, which are equal to infinite products. It would be interesting to generalize the integral-product identities and figure out the boundary confining dualities with line operators.  
    
    \item  It would be nice to give analytic proofs of our formulas for arbitrary $SU(N)$ gauge group. 
    For higher rank gauge groups, they will be obtained from a certain multivariable extension of the Jacobi triple product identity. 
    We also expect that they can be addressed by using the Fermi-gas method based on the determinant formula. 

    \item The grand canonical ensemble of one-point functions of charged Wilson lines for $SU(N)$ pure Chern-Simons involves the mock modular Appell-Lerch sums. What is the physical meaning of the shadow? Is there an interpretation in terms of a dual gravity theory? This might be possible to understand from a brane configuration giving these 3d Chern-Simons theories with boundaries but note that it is not straightforward to construct such a theory with gauge group $SU(N)$ rather than $U(N)$ in 3d.

    \item We have presented several $q$-difference equations relating half-indices to half-indices with Wilson lines. It would be good to get a better understanding of the origin and interpretation of these $q$-difference equations.

    \item There are different ways to construct the $SU(N)_k$ theories with Neumann boundary conditions for the vector multiplet. Cancellation of the gauge anomaly can be achieved with different combinations of fundamental and anti-fundamental 3d chirals an 2d fundamental Fermis. It would be interesting to understand the dual descriptions in general.
    
\end{itemize}

%%%%%%%%%%%%%%%%%%%%%%%%%%%%%%%%%%
%%%%%%%%%%%%%%%%%%%%%%%%%%%%%%%%%%
\section{CS theory with Wilson lines and boundary}
\label{sec_CS_suN}
%%%%%%%%%%%%%%%%%%%%%%%%%%%%%%%%%%
%%%%%%%%%%%%%%%%%%%%%%%%%%%%%%%%%%
%\textcolor{red}{[NOTE]: The configuration should be refined in the presence of the additional Neumann for chiral in the case with $k=-N-3/2,\cdots$, say by introducing additional parameter labeling the number of such chiral?}

In this section we briefly review relevant features of $\mathcal{N} = 2$ supersymmetric $SU(N)$ Chern-Simons theories with boundary. In addition to the vector multiplet we consider fundamental and anti-fundamental chirals as well as the possibility of fundamental 2d chiral or Fermi multiplets. 
We focus on the case of Neumann boundary conditions for the vector multiplet and present the anomaly polynomial. 
Cancellation of the gauge anomaly determines the Chern-Simons coupling. We then discuss the half-index with the addition of Wilson lines.

%%%%%%%%%%%%%%%%%%%%%%%%%%%%%%%%%%
\subsection{Anomalies}
\label{sec_anomalies}
%%%%%%%%%%%%%%%%%%%%%%%%%%%%%%%%%%
For pure $SU(N)_k$ theory at level $k$ with Neumann boundary conditions for the vector multiplet, $N_f$ fundamental and $N_a$ antifundamental 3d chiral multiplets $Q_I$ and $\overline{Q}_{\alpha}$ with Neumann boundary conditions, and $M$ fundamental 2d Fermi multiplets we have anomaly polynomial \cite{Dimofte:2017tpi}
\begin{align}
    \Acal & = \underbrace{k\Tr(s^2)}_{\textrm{CS}}
 + \underbrace{N\Tr(s^2) + \frac{N^2 - 1}{2}r^2}_{\textrm{VM}, \; \Ncal}
 - \underbrace{\left( \frac{N_f}{2} \Tr(s^2) + \frac{N}{2} \Tr(x^2) + \frac{N N_f}{2}(a-r)^2 \right)}_{Q_I, \; N}
 \nonumber \\
 & - \underbrace{\left( \frac{N_a}{2} \Tr(s^2) + \frac{N}{2} \Tr(\tilde{x}^2) + \frac{N N_a}{2}(b-r)^2 \right)}_{\overline{Q}_{\alpha}, \; N}
 + \underbrace{M\Tr(s^2) + N\Tr(\tilde{s}^2)}_{\textrm{Fermi}}
 \; .
\end{align}
Here $s$ represents the $SU(N)$ gauge field strength, $\tilde{s}$ the global $U(M)$ field strength, $r$ the $U(1)$ R-charge, $x$ the global $SU(N_f)$ flavor symmetry with $U(1)_a$ axial symmetry, and  $\tilde{x}$ the global $SU(N_a)$ flavor symmetry with $U(1)_b$ axial symmetry.

We see that in order to cancel the gauge anomaly we must have Chern-Simons level
\begin{align}
\label{suN_anom_CS}
    k & = -\left( N - \frac{N_f + N_a}{2} + M \right) \; .
\end{align}
Of course, for $SU(2)$ antifundamentals are the same as fundamentals so we can always set $N_a = 0$.

Following similar arguments given in \cite{Dimofte:2017tpi} we expect a dual $U(M)_{-k, -k - M}$ theory with Dirichlet boundary conditions for the vector multiplet, and $N_f$ fundamental and $N_a$ antifundamental 3d chiral multiplets $\tilde{Q}_I$ and $\overline{\tilde{Q}}_{\alpha}$ with Dirichlet boundary conditions. In particular the contribution to the $U(M)$ anomaly matches above with
\begin{align}
\Acal & \sim \underbrace{-k\Tr(\tilde{s}^2) - (\Tr(\tilde{s}))^2}_{\textrm{CS}}
 \underbrace{-M\Tr(\tilde{s}^2) + (\Tr(\tilde{s}))^2}_{\textrm{VM}}
 + \underbrace{\frac{N_f}{2} \Tr(\tilde{s}^2)}_{Q_I, \; D}
 + \underbrace{\frac{N_a}{2} \Tr(\tilde{s}^2)}_{\overline{Q}_I, \; D}
\end{align}

%We have the anomaly polynomial
%\begin{align}
%    \Acal & = \underbrace{-k\Tr(\tilde{s}^2) + \frac{N}{M}(\Tr(\tilde{s}))^2}_{\textrm{CS}}
% - \underbrace{M\Tr(\tilde{s}^2) + (\Tr(\tilde{s}))^2 - \frac{M^2 - 1}{2}r^2}_{\textrm{VM}, \; \Dcal}
% \nonumber \\
% & + \underbrace{\left( \frac{N_f}{2} \Tr(\tilde{s}^2) + N_f\Tr(\tilde{s})(a-r) + \frac{M}{2} \Tr(x^2) + \frac{M N_f}{2}(a-r)^2 \right)}_{Q_I, \; D}
% \nonumber \\
% & + \underbrace{\left( \frac{N_a}{2} \Tr(\tilde{s}^2) + N_a\Tr(\tilde{s})(b-r) + \frac{M}{2} \Tr(\tilde{x}^2) + \frac{M N_a}{2}(b-r)^2 \right)}_{\overline{Q}_{\alpha}, \; D}
% \; .
%\end{align}

%%%%%%%%%%%%%%%%%%%%%%%%%%%%%%%%%%
\subsection{Wilson line half-indices}
\label{sec_wilsonline}
%%%%%%%%%%%%%%%%%%%%%%%%%%%%%%%%%%
As shown in \cite{Dimofte:2017tpi} the half-index for the theories described here are built from the vector multiplet with Neumann boundary conditions contribution
\begin{align}
    \frac{(q)_{\infty}^{N-1}}{N!} \oint \left( \prod_{i = 1}^{N-1} \frac{ds_i}{2\pi i s_i} \right) \prod_{1 \le i < j \le N} (s_i^{\pm} s_j^{\mp}; q)_{\infty}, 
\end{align}
where the integral ensures gauge invariance and for $SU(N)$ we have the restriction on the gauge fugacities $\prod_{i = 1}^N s_i = 1$.

Including $N_f$ fundamental 3d chirals with R-charge $r$ and Neumann boundary conditions we should include a factor 
\begin{align}
    \prod_{\alpha = 1}^{N_f} \prod_{i = 1}^N \frac{1}{(q^{\frac{r}{2}} s_i x_{\alpha}; q)_{\infty}}, 
\end{align}
where $x_{\alpha}$ are $U(N_f)$ flavor symmetry fugacities (which are often split into a $U(1)$ axial and $SU(N_f)$ flavor symmetry fugacities). Anti-fundamental chirals give a similar contribution with $s_i \rightarrow s_i^{-1}$ and $x_{\alpha} \rightarrow \tilde{x}_I$ for the $U(N_a)$ global symmetry.

Instead, Dirichlet boundary conditions for a fundamental would give contribution
\begin{align}
    \prod_{\alpha = 1}^{N_f} \prod_{i = 1}^N (q^{1 - \frac{r}{2}} s_i^{-1} x_{\alpha}^{-1}; q)_{\infty} \; .
\end{align}

On the boundary we can have 2d Fermi or chiral multiplets. In the fundamental representation these Fermi multiplets with R-charge $0$ give the same contribution as the combination of a pair of fundamental and anti-fundamental 3d chirals with R-charge $r = 1$ and Dirichlet boundary conditions up to an identification of flavor fugacities. In particular, for $M$ fundamental 2d Fermi multiplets we have contribution
\begin{align}
    \prod_{\alpha = 1}^{M} \prod_{i = 1}^N (q^{\frac{1}{2}} s_i^{\pm} x_{\alpha}^{\pm}; q)_{\infty} \; .
\end{align}
where the Fermis are in the fundamental representation of a global $U(M)$ symmetry. 

To make the above statement precise, this is equivalent to the case of $N_f = N_a = M$ where the 3d chirals have Dirichlet boundary conditions and we have specialized the $U(N_a)$ flavor fugacities $\tilde{x}_{\alpha} \rightarrow x_{\alpha}^{-1}$.

Also, note that with further identification of flavor fugacities, a 2d fundamental Fermi together with a fundamental 3d chiral with Neumann boundary conditions and R-charge $1$ gives the same contribution as a 3d fundamental chiral with Dirichlet boundary conditions and R-charge $1$. Specifically, there is a partial cancellation in the half-index contribution if the 3d chiral and 2d Fermi both have the same flavor fugacity $x_{\alpha}$.
%\textcolor{red}{Add some more detail and more general explanation.}

We can include line operators ending at a point on the boundary. In particular we consider Wilson lines. If the Wilson line is in a representation $R$ of the gauge group the half-index will be modified to count BPS states in the conjugate representation $\overline{R}$, so that the total configuration including the Wilson line is gauge invariant This is simply implemented \cite{Dimofte:2017tpi} by including a factor $\Tr_R s$ in the half-index. For example if we consider a Wilson line in the fundamental representation of the gauge group we should include a factor $\sum_{i = 1}^N s_i$. More generally, the irreducible representation (irrep) $R$ of $SU(N)$ is labeled by the Young diagram $\lambda$, 
for which we should introduce the Schur function $s_{\lambda}(s)$. 
We denote by $W_{\lambda}$ the Wilson line in the irrep labeled by $\lambda$. 
Alternatively, the power sum symmetric function
\begin{align}
p_n(s)&=\sum_{i=1}^{N}s_i^n
\end{align}
associated with conjugacy classes of the symmetric group can be introduced. 
We will call the Wilson line operator labeled by the power sum symmetric function $p_n(s)$ of degree $n$ the \textit{charge-$n$ Wilson line} $W_n$ and we will denote the half-index in the presence of such a Wilson line by $\langle W_n \rangle$.

Instead, with Dirichlet boundary conditions for a $U(M)$ gauge group we have vector multiplet contribution
\begin{align}
    \frac{1}{(q)_{\infty}^N} \sum_{m_i \in \Zb} \frac{q^{\frac{k}{2}\sum_i m_i^2 + \frac{\gamma}{2}(\sum_i m_i)^2} (\prod_i u_i^{km_i + \gamma \sum_j m_j})}{\prod_{i \ne j} (q^{1 + m_i - m_j} u_i u_j^{-1}; q)_{\infty}}, 
\end{align}
and for other factors in the half-index we shift the gauge fugacities $u_i \rightarrow q^{m_i} u_i$, with the gauge fugacities $u_i$ all independent unlike for special unitary groups. Here $k$ is the effective Chern-Simons level and $\gamma$ encodes a possibly different effective Chern-Simons level for the diagonal $U(1)$ subgroup, in the sense that the overall anomaly polynomial has contribution
$k\Tr(\tilde{s}^2) + \gamma(\Tr(\tilde{s}))^2$.

%%%%%%%%%%%%%%%%%%%%%%%%%%%%%%%%%%
\subsection{$q$-difference equations}
\label{sec_qdiff}
%%%%%%%%%%%%%%%%%%%%%%%%%%%%%%%%%%

Half-indices can satisfy $q$-difference equations which are constructed using operators which scale fugacities by powers of $q$, e.g.\ $x \to qx$.
This provides one way to prove identities of half-indices, e.g.\ if two half-indices obey the same first order $q$-difference equation and can be shown to match for specific values of global fugacities. Several details and examples in the context of half-indices are given in \cite{Dimofte:2017tpi}.

We will also see examples where half indices with line operators are related to the standard half-index without a line operator. In the case of Wilson lines in $SU(N)$ theories with Neumann boundary conditions for the vector multiplet this involves shifts of global fugacities by fractional powers of $q$, in particular by multiples of $q^{1/N}$. For example, in the case of $SU(2)$, if we label a half-index as $\mathbb{II}(x;q)$ and with a line operator as $\langle W \rangle (x;q)$ where $x$ is a global fugacity, we find relations such as
\begin{align}
\label{qdiff_line_example}
\mathbb{II}(q^{1/2}x;q)
&=-q^{-1/2}x^{-1}\langle W \rangle(x;q) \; .
\end{align}
These fractional powers of $q$ can be understood in the context of non-Abelian line operators carrying fractional charge and spin which can be understood in terms of ``Cheshire charge/spin''
\cite{Alford:1990ur, Preskill:1990bm, Bucher:1991bc, Alford:1992yx, Brekke:1992ft}.
In the case of $SU(N)$ Wilson lines we find power of $q^{1/N}$ corresponding fractional charges and spins in units of $1/N$. However, we have not been able to understand the precise structure of the $q$-difference equations and we leave that to future work.

%%%%%%%%%%%%%%%%%%%%%%%%%%%%%%%%%%
\subsection{Special functions and $q$-series identities}
\label{sec_SpecFns}
%%%%%%%%%%%%%%%%%%%%%%%%%%%%%%%%%%

To derive some of the exact results we use the Jacobi triple product identity which can be written in the form
\begin{align}
(q)_{\infty} (x^{-1}; q)_{\infty} (qx; q)_{\infty} & = \sum_{m = -\infty}^{\infty} (-1)^m q^{m(m+1)/2} x^m \; .
\end{align}
This is equivalent to
\begin{align}
\label{JacobiTripleProduct}
(q)_{\infty} (x^{\pm}; q)_{\infty} & = (1-x) \sum_{m = -\infty}^{\infty} (-1)^m q^{m(m+1)/2} x^m
\end{align}
and this form is used in several derivations to replace products of $q$-Pochhammer symbols with sums. In the case of Neumann boundary conditions for the vector multiplet, in general this enables the half-index to be expressed as an integral of multiple sums. The integrals can then be evaluated simply using the Cauchy residue theorem and the resulting sum can be interpreted as a half-index with Dirichlet boundary conditions for a dual vector multiplet.

%%%%%%%%%%%%%%%%%%%%%%%%%%%%%%%%%%
\subsection{Rogers-Ramanujan functions}
\label{sec_RRfunction}
%%%%%%%%%%%%%%%%%%%%%%%%%%%%%%%%%%
One intriguing result is that 
the following pair of functions occurs in the calculation of the half-index and the one-point function of the Wilson line in the fundamental representation:  
\begin{align}
\label{RR_identity1}
G(q)&=\sum_{n=0}^{\infty}
\frac{q^{n^2}}{(1-q)(1-q^2)\cdots (1-q^n)}
\nonumber\\
&
=\prod_{n=1}^{\infty}\frac{1}{(1-q^{5n-1})(1-q^{5n-4})}
=\frac{f(-q^2,-q^3)}{f(-q)}
, \\
\label{RR_identity2}
H(q)&=\sum_{n=0}^{\infty}
\frac{q^{n(n+1)}}{(1-q)(1-q^2)\cdots (1-q^n)}
\nonumber\\
&=\prod_{n=1}^{\infty}\frac{1}{(1-q^{5n-2})(1-q^{5n-3})}
=\frac{f(-q,-q^4)}{f(-q)}
\end{align}
where $f(a,b)$ and $f(-q)$ are general Ramanujan's theta functions (\ref{r_theta}) and (\ref{r_theta_f}). 
The product-series equalities (\ref{RR_identity1}) and (\ref{RR_identity2}) are known as the Rogers-Ramanujan identities (e.g. see \cite{MR1634067}). 
These identities are of the greatest significance in the theory of partitions and number theory \cite{MR1634067}. 
The function (\ref{RR_identity1}) is the generating function for partitions of $n$ into parts with minimal difference $2$ with all parts greater than $0$ 
or equivalently that for partitions of $n$ of the form $5k+1$ and $5k+4$. 
The function (\ref{RR_identity2}) is the generating function for partitions of $n$ into parts with minimal difference $2$ with all parts greater than $1$ 
or equivalently that for partitions of $n$ of the form $5k+2$ and $5k+3$. 

In section \ref{sec_su2_km52}, we find novel physical realizations of the functions associated with the Rogers-Ramanujan identities. 
It follows that the first function (\ref{RR_identity1}) and the second function (\ref{RR_identity2}) 
are respectively identified with the half-index and the one-point function of the fundamental Wilson line 
for $SU(2)_{-5/2}$ CS theory coupled with a single chiral multiplet in the fundamental representation 
where the vector multiplet obeys the Neumann boundary condition and the chiral multiplet the Dirichlet boundary condition. 
Consequently, we get integral representations of the functions (\ref{RR_identity1}) and (\ref{RR_identity2})
\begin{align}
G(q)&=\frac{(q;q)_{\infty}}{2} \oint \frac{ds}{2\pi is} (s^{\pm 2};q)_{\infty}(q^{\frac12}s^{\pm};q)_{\infty}^2, \\
H(q)&=-q^{-1/2}\frac{(q;q)_{\infty}}{2} \oint \frac{ds}{2\pi is} (s^{\pm 2};q)_{\infty}(q^{\frac12}s^{\pm};q)_{\infty}^2(s+s^{-1}). 
\end{align}
With flavor fugacity $x$ we find that
\begin{align}
\label{gRR1}
\sum_{n=0}^{\infty} \frac{q^{n^2}x^{2n}}{(1-q)(1-q^2)\cdots (1-q^n)}
 & = \frac{(q;q)_{\infty}}{2} \oint \frac{ds}{2\pi is} (s^{\pm 2};q)_{\infty}(q^{\frac12}s^{\pm}x;q)_{\infty}^2, \\
\label{gRR2}
\sum_{n=0}^{\infty}
\frac{q^{n(n+1)}x^{2n+1}}{(1-q)(1-q^2)\cdots (1-q^n)}& = -q^{-1/2}\frac{(q;q)_{\infty}}{2} \oint \frac{ds}{2\pi is} (s^{\pm 2};q)_{\infty}(q^{\frac12}s^{\pm}x;q)_{\infty}^2 (s+s^{-1}). 
\end{align}
The L.H.S. of (\ref{gRR1}) and (\ref{gRR2}) are sometimes called Rogers-Ramanujan functions \cite{MR1117903}. 
More general integral-series equalities and their physical interpretatoins are found 
in the study of higher rank CS theory theories. 

%%%%%%%%%%%%%%%%%%%%%%%%%%%%%%%%%%
%%%%%%%%%%%%%%%%%%%%%%%%%%%%%%%%%%
\section{$SU(2)$}
\label{sec_su2}
%%%%%%%%%%%%%%%%%%%%%%%%%%%%%%%%%%
%%%%%%%%%%%%%%%%%%%%%%%%%%%%%%%%%%
In this section we evaluate the line defect half-indices for the 3d $\mathcal{N}=2$ supersymmetric 
$SU(2)$ Chern-Simons theory with Wilson line operators and Neumann boundary conditions for the vector multiplet. We also include $N_f$ fundamental chirals with Neumann boundary conditions and $M$ fundamental 2d Fermis. In fact, we will focus on the cases $N_f = 0$ for integer Chern-Simons level and $N_f = 1$ for half-integer level.
Recalling (\ref{suN_anom_CS}) we see that the Chern-Simons level is
\begin{align}
\label{su2_anom_CS}
    k & = -\left( 2 - \frac{N_f}{2} + M \right) \; .
\end{align}

For gauge group $SU(2)$ we have gauge fugacities $s_1$ and $s_2$ with a constraint $s_1 s_2 = 1$. 
This can be expressed in terms of a single independent gauge fugacity $s$ where $s_1 = s$ and $s_2 = s^{-1}$. 
With Neumann boundary conditions for the vector multiplet, 
the half index will contain a contour integral w.r.t.\ $s$ of the vector multiplet contribution 
multiplied by finite or infinite power series in $s$ to include the matter and/or line operator contributions.

%%%%%%%%%%%%%%%%%%%%%%%%%%%%%%%%%%
%%%%%%%%%%%%%%%%%%%%%%%%%%%%%%%%%%
\subsection{$k = -2$}
\label{sec_su2_km2}
%%%%%%%%%%%%%%%%%%%%%%%%%%%%%%%%%%
%%%%%%%%%%%%%%%%%%%%%%%%%%%%%%%%%%
When $N_f=0$ and $M=0$, we find the consistent $\mathcal{N}=(0,2)$ BPS boundary conditions 
for 3d $\mathcal{N}=2$ $SU(2)$ pure Chern-Simons theory with level $k=-2$. 

Unlike the previous cases, the theory has no matter field, 
so there is no non-trivial gauge invariant BPS local operator at the boundary 
preserving $\mathcal{N}=(0,2)$ supersymmetry without any insertion of line defect. 
In fact, the half-index is trivial
\begin{align}
\mathbb{II}_{\mathcal{N}}^{SU(2)_{-2}}&=\frac{(q;q)_{\infty}}{2}\oint \frac{ds}{2\pi is}(s^{\pm2};q)_{\infty}=1.
\end{align}

%%%%%%%%%%%%%%%%%%%%%%%%%%%%%%%%%%
\subsubsection{One-point function}
\label{sec_1pt_su2km2}
%%%%%%%%%%%%%%%%%%%%%%%%%%%%%%%%%%
Insertion of the Wilson line in the fundamental representation ending on the boundary is not allowed in this case 
since there is no candidate that can cancel the gauge anomaly while preserving $\mathcal{N}=(0,2)$ supersymmetry. 
In fact, it can be shown that the one-point function of the Wilson line in the fundamental representation vanishes
\begin{align}
\langle W_1\rangle^{SU(2)_{-2}}
&=\frac{(q;q)_{\infty}}{2}\oint 
\frac{ds}{2\pi is}(s^{\pm2};q)_{\infty} (s+s^{-1})=0. 
\end{align}

Non-trivial one-point functions appear for the Wilson lines transforming in more general representations.  
We find that the one-point function of the Wilson line $W_{(k)}$ in the rank-$k$ symmetric representation for $k$ even is given by
\begin{align}
\langle W_{(k)} \rangle^{SU(2)_{-2}} & = (-1)^{k/2} q^{k(k+2)/8} \; .
\end{align}
while the one-point function vanishes for $k$ odd. 

This can be understood by noting that the half-index in the presence of a Wilson line in some representation counts the BPS operators which can be combined with the Wilson line representation to form a gauge singlet. In this case that means the operators in representations which can be combined with the rank-$k$ symmetric representation to form a singlet. Now, the only boundary BPS operators are the gaugino $\lambda_{-}$ and its derivatives $D_z^n \lambda_{-}$ which are in the adjoint (or rank-$2$ symmetric) representation of $SU(2)$ and contribute fugacity $-q^{n+1}$ in addition to gauge fugacities. To contract the gauge indices of the rank-$k$ symmetric representation we must take at least $k/2$ such operators, noting that for $SU(2)$ anti-fundamental indices are equivalent to fundamental indices. Since the gauginos (and their derivatives) are fermionic these operators must all be distinct so the minimal choice is each $n$ in the range $0 \le n \le k/2 - 1$, giving fugacity $\prod_{n = 0}^{k/2 - 1} -q^{n + 1} = (-1)^{k/2} q^{\sum_{n = 0}^{k/2 - 1} (n + 1)} = (-1)^{k/2} q^{k(k+2)/8}$.
We might expect higher order contributions (by including more derivatives) but then there are other operators with the same fugacity where we replace some derivatives by gauginos. This leads to cancellations since each gaugino contributes a factor $-1$, and the result is that there is a total cancellation of such terms except for the minimal case where there is only a single operator to contribute.

Let us give an analytic proof by directly calculating the half-index. 
All one-point functions can be obtained if we can evaluate the matrix integral with a monomial contribution
\begin{align}
I_{\alpha} & = (q)_{\infty} \oint \frac{ds}{2 \pi i s} (s^{\pm 2}; q)_{\infty} s^{\alpha}
\end{align}
where $\alpha \in \Zb$.

We first use the Jacobi triple product identity \eqref{JacobiTripleProduct} which gives
\begin{align}
I_{\alpha} & = \oint \frac{ds}{2 \pi i s} \sum_{m = -\infty}^{\infty} (-1)^m q^{m(m+1)/2} (1 - s^2) s^{\alpha + 2m} \; .
\end{align}
The contour integral picks out only the terms with $m = -\alpha/2$ or $m = -\alpha/2 - 1$ so we see that the result is zero if $\alpha$ is odd. Assuming $\alpha$ is even we find
\begin{align}
    I_{\alpha} & = (-1)^{\alpha/2} \left( q^{\alpha(\alpha - 2)/8} + q^{\alpha(\alpha + 2)/8} \right) \; .
\end{align}

So, for example if $n = 2k$ is even the half-index with a charge-$n$ Wilson line is
\begin{align}
    \langle W_n \rangle^{SU(2)_{-2}} 
    &= \frac{(q)_{\infty}}{2} \oint \frac{ds}{2 \pi i s} (s^{\pm 2}; q)_{\infty} \left( s^n + s^{-n} \right) 
   \nonumber\\
    & = \frac{(-1)^k}{2} \left( q^{k(k - 1)/2} + q^{k(k + 1)/2} \right), 
\end{align}
while this integral vanishes for odd $n$. 
Accordingly, we can also evaluate 
the one-point function for rank-$k$ symmetric representation Wilson lines $W_{(k)}$ for $k$ even
\begin{align}
\langle W_{(k)} \rangle^{SU(2)_{-2}} & = \langle \sum_{\alpha = 0}^k s_1^{\alpha} s_2^{k - \alpha} \rangle
= \langle \sum_{\alpha = 0}^k s^{2\alpha - k} \rangle
= 1 + \sum_{\beta = 1}^{k/2} \langle s^{2\beta} + s^{-2\beta} \rangle
\nonumber \\
& = 1 + \sum_{\beta = 1}^{k/2} (-1)^{\beta} \left( q^{\beta(\beta - 1)/2} + q^{\beta(\beta + 1)/2} \right)
\nonumber \\
& = (-1)^{k/2} q^{k(k+2)/8} \label{su2_1pt_sym} \; .
\end{align}

%%%%%%%%%%%%%%%%%%%%%%%%%%%%%%%%%%
\subsubsection{Grand canonical one-point function}
\label{sec_1pt_su2G}
%%%%%%%%%%%%%%%%%%%%%%%%%%%%%%%%%%
Now, let us consider the case where $n = 2k$ and calculate the grand canonical ensemble by summing over the charge $n$. 
%or equivalently over the rank $k$ of the symmetric representation. 
We obtain the grand canonical one-point function of the charged Wilson line 
\begin{align}
\langle W_{\textrm{charged}}\rangle^{SU(2)_{-2}}
&=\sum_{k \in \Zb} \langle W_{2k} \rangle^{SU(2)_{-2}} \Lambda^{k}
\nonumber\\
&= -i \Lambda^{1/2} \sum_{m = 0}^{1} q^{m/2 - 1/8} \vartheta_1(\lambda + m\tau; \tau)
\end{align}
where $\Lambda = \exp(2\pi i \lambda)$ and the Jacobi theta series is defined in \eqref{theta1}
%\begin{align}
%    \vartheta_!(v; \tau) = & \sum_{\rho \in \Zb + 1/2} (-1)^{\rho} V^{\rho} q^{\rho^2/2} \; .
%\end{align}
%where $V = \exp(2\pi i v)$.

%%%%%%%%%%%%%%%%%%%%%%%%%%%%%%%%%%
\subsubsection{Two-point function}
\label{sec_2pt_su2k2sym}
%%%%%%%%%%%%%%%%%%%%%%%%%%%%%%%%%%
We can also exactly evaluate the two-point function $\langle W_{(k)} W_{(l)} \rangle$ of the Wilson lines in the symmetric representations which vanishes if $l-k$ is odd. 
If $l-k$ is even define integers $m = (k+l)/2$ and $\delta = (l - k)/2$. 
We then have
\begin{align}
\label{su2_2pt_sym}
\langle W_{(k)} W_{(l)} \rangle^{SU(2)_{-2}} 
& = \sum_{\alpha = 0}^k \sum_{\beta = 0}^l \langle s^{2\alpha - k} s^{2\beta - l} \rangle
\nonumber \\
& = \frac{1}{2} \sum_{\alpha = 0}^k \sum_{\beta = 0}^l (-1)^{\alpha + \beta - m} \left(q^{(\alpha + \beta - m) (\alpha + \beta - m - 1)/2} + q^{(\alpha + \beta - m) (\alpha + \beta - m + 1)/2} \right)
\nonumber \\
& = \frac{1}{2} \sum_{\alpha = 0}^k \left( (-1)^{\alpha - m} q^{(\alpha - m) (\alpha - m - 1)/2} + (-1)^{\alpha + l - m} q^{(\alpha + l - m) (\alpha + l - m + 1)/2} \right)
\nonumber \\
& = \sum_{\alpha = 0}^k (-1)^{\alpha - m} q^{(\alpha - m) (\alpha - m - 1)/2}
\nonumber \\
& = \sum_{\alpha = 0}^k (-1)^{\alpha + \delta} q^{(\alpha + \delta) (\alpha + \delta + 1)/2}
= \sum_{\alpha = \delta}^{k + \delta} (-1)^{\alpha} q^{\alpha (\alpha + 1)/2}. 
\end{align}

%\textcolor{red}{[NOTE] Physical interpretation}
Here we can interpret this result in terms of the expansion of the product of rank-$k$ and rank-$l$ symmetric representations into a sum of symmetric representations and then using the result for the one-point function \eqref{su2_1pt_sym}.

When $k=l$, the two-point function (\ref{su2_2pt_sym}) becomes
\begin{align}
\langle W_{(k)} W_{(k)} \rangle^{SU(2)_{-2}} 
&=\sum_{\alpha=0}^{k}(-1)^{\alpha} q^{\frac{\alpha(\alpha+1)}{2}}. 
\end{align}
In the large symmetric representation limit $k\rightarrow \infty$, it agrees with the false theta function introduced by Rogers \cite{rogers1917two}
\begin{align}
\label{su2_2pt_largesym}
\langle W_{(\infty)} W_{(\infty)} \rangle^{SU(2)_{-2}} 
&=f(q)=\sum_{\alpha=0}^{\infty}(-1)^{\alpha}q^{\frac{\alpha(\alpha+1)}{2}}. 
\end{align}

%%%%%%%%%%%%%%%%%%%%%%%%%%%%%%%%%%
%%%%%%%%%%%%%%%%%%%%%%%%%%%%%%%%%%
\subsection{$k= - 5/2$}
\label{sec_su2_km52}
%%%%%%%%%%%%%%%%%%%%%%%%%%%%%%%%%%
%%%%%%%%%%%%%%%%%%%%%%%%%%%%%%%%%%
Taking $N_f = 1$ and $M = 1$ gives $k = -\frac{5}{2}$. Alternatively we could take a single fundamental chiral with Dirichlet boundary conditions with no Fermis.
Specifically, the former case with identification of the Fermi $U(1)$ flavor fugacity $u = x$ and the 3d chiral $U(1)$ flavor fugacity $x^{-1}$ gives the same half-index as 3d $\mathcal{N}=2$ $SU(2)$ Chern-Simons-matter theory 
with level $k=-5/2$ and a single fundamental chiral multiplet with Dirichlet boundary condition. 
The half-index is 
\begin{align}
\label{h_su2k5/2}
\mathbb{II}_{\mathcal{N}}^{SU(2)_{-5/2}}(x;q)
&=
\frac{(q;q)_{\infty}}{2} \oint \frac{ds}{2 \pi i s} (s^{\pm 2}; q)_{\infty} 
(q^{\frac12}sx;q)_{\infty}
(q^{\frac12}s^{-1}x;q)_{\infty}. 
\end{align}
We can expand the half-index as
\begin{align}
&1+x^2q+x^2q^2+x^2q^3+(x^2+x^4)q^4+(x^2+x^4)q^5
\nonumber\\
&+(x^2+2x^4)q^6+(x^2+2x^4)q^7+(x^2+3x^4)q^8+(x^2+3x^4+x^6)q^9+\cdots
\end{align}
This result is easily understood as the counting of the gauge-invariant operators formed from the fundamental chiral with Dirichlet boundary conditions and its derivatives, taking into account the fact that gauge invariants can only be formed by pairing of these fermions and the coefficients arise from different distributions of the derivatives taking into account the antisymmetric properties. The interpretation in terms of a 2d Fermi and 3d chiral with Neumann boundary conditions is more complicated.

We find that the half-index agrees with the Rogers-Ramanujan function \cite{MR1117903}
\begin{align}
\label{RR_sum}
\sum_{n=0}^{\infty} \frac{q^{n^2}x^{2n}}{(1-q)(1-q^2)\cdots (1-q^n)}
 & = \sum_{n = -\infty}^{\infty} \frac{q^{n^2} x^{2n} (q^{n+1}; q)_{\infty}}{(q)_{\infty}} \; . 
\end{align}

Taking the interpretation as $SU(2)_{-5/2}$ with $N_f = 1$ and $M = 1$ where the fundamental chiral has R-charge $1$, we expect a dual description as a $U(1)_{3/2}$ theory with Dirichlet boundary conditions for the vector multiplet and a charged 3d chiral with R-charge $0$ and Dirichlet boundary conditions. This would give the half-index
\begin{align}
\label{h_u1k3/2_D}
\mathbb{II}_{\mathcal{D}}^{U(1)_{3/2}}(x;q)
&=
\frac{1}{(q)_{\infty}} \sum_{n \in \Zb} q^{n^2} u^{2n} (q^{1 + n} u x^{-1}; q)_{\infty}
\end{align}
which matches \eqref{RR_sum} in the case $u = x$.

In the unflavored limit $x\rightarrow 1$, we obtain (\ref{RR_identity1}), the first functions associated with the Rogers-Ramanujan identities! 
Namely, we find
\begin{align}
&\mathbb{II}_{\mathcal{N}}^{SU(2)_{-5/2}}(x=1;q)
=G(q)
\nonumber \\
 & = 
\frac{1}{(q)_{\infty}} \sum_{m \in \Zb} (-1)^m q^{\frac{5}{2}m^2 + \frac{1}{2}m}
=\frac{1}{(q)_{\infty}}\sum_{m\in \mathbb{Z}}(q^{10m^2+m}-q^{10m^2+9m+2}). 
\end{align}
We observe that 
the associated boundary vertex operator algebra (VOA) arise 
from the Virasoro minimal model $\mathcal{M}(2,5)$ 
by observing that it agrees with the character $\chi_{1,2}^{2,5}(q)$ where
\begin{align}
\chi_{r,s}^{p,p'}(q)
&=\frac{1}{(q;q)_{\infty}}
\sum_{m\in \mathbb{Z}}(q^{m^2pp'+m(p'r-ps)}-q^{(mp+r)(mp'+s)})
\end{align}
is the normalized Virasoro character \cite{MR781391} of the minimal model $\mathcal{M}(p,p')$ \footnote{Here $1<p<p'$ with $p$ and $p'$ being relatively coprime. }
with conformal dimension
\begin{align}
h_{r,s}^{p,p'}&=\frac{(rp'-sp)^2-(p-p')^2}{4pp'}, \qquad 
1\le r\le p-1, 1\le s\le p'-1
\end{align}
and central charge 
\begin{align}
c(p,p')&=1-\frac{6(p-p')^2}{pp'}. 
\end{align}
Including the normalization factor $q^{h_{r,s}^{p,p'}-c(p,p')/24}$, we get 
\begin{align}
\widetilde{G}(q)=q^{-1/60}G(q). 
\end{align}

%modular
Also we observe that the half-index has a nice behavior under the modular transformation. 
We have \cite{MR3774207}
\begin{align}
\widetilde{G}\left(\frac{a\tau+b}{c\tau+d}\right)
&=e^{2\pi i\alpha(a,b,c)/60}\widetilde{G}(\tau)
\end{align}
where 
\begin{align}
\alpha(a,b,c)&=a(9-b+c)-9
\end{align}
and $q=e^{2\pi i\tau}$. 

%%%%%%%%%%%%%%%%%%%%%%%%%%%%%%%%%%
\subsubsection{One-point function}
\label{sec_1pt_su2km5/2}
%%%%%%%%%%%%%%%%%%%%%%%%%%%%%%%%%%
The one-point function of the Wilson line in the fundamental representation is evaluated as
\begin{align}
\langle W_1\rangle^{SU(2)_{-5/2}}(x;q)
&= \frac{(q;q)_{\infty}}{2} \oint \frac{ds}{2 \pi i s} (s^{\pm 2}; q)_{\infty} 
(q^{\frac12}sx;q)_{\infty}
(q^{\frac12}s^{-1}x;q)_{\infty}
(s+s^{-1}). 
\end{align}
It can be expanded as
\begin{align}
&-xq^{1/2}-x^3q^{5/2}-x^3q^{7/2}-x^3q^{9/2}-x^3q^{11/2}-(x^3+x^5)q^{15/2}
\nonumber\\
&-(x^3+2x^5)q^{17/2}-(x^3+2x^5)q^{19/2}-\cdots. 
\end{align}
%\textcolor{red}{[NOTE] Physical interpretation}
As for the half-index without a line operator,
this result is easily understood as the counting of the gauge-invariant operators formed from the fundamental chiral and its derivatives, taking into account the fact that gauge invariants can only be formed by pairing of these fermions. Now, due to the fundamental Wilson line we must count operators in the fundamental representation. This means pairing a chiral (possibly with derivatives) with the Wilson line and then adding other pairs of chirals, and distributing derivatives so that the product does not vanish due to antisymmetric properties.

We find that the one-point function is given by 
\begin{align}
\label{RR_sum2}
\langle W_1\rangle^{SU(2)_{-5/2}}(x;q) & = 
-q^{1/2}
\sum_{n=0}^{\infty}
\frac{q^{n(n+1)}x^{2n+1}}{(1-q)(1-q^2)\cdots (1-q^n)}. 
\end{align}

%different equation
Note that we have the $q$-difference equation 
\begin{align}
\label{qdiff_su2k5/2}
\mathbb{II}_{\mathcal{N}}^{SU(2)_{-5/2}}(q^{1/2}x;q)
&=-q^{-1/2}x^{-1}\langle W_{1}\rangle^{SU(2)_{-5/2}}(x;q) \; .
\end{align}

In the unflavored limit $x\rightarrow 1$, 
the one-point function coincides with 
\begin{align}
-q^{1/2}H(q)
 = 
\frac{1}{(q;q)_{\infty}} \sum_{m \in \Zb} (-1)^{m + 1} q^{\frac{5}{2}m^2 - \frac{3}{2}m + \frac{1}{2}}
\end{align}
where $H(q)$ is the second function (\ref{RR_identity2}) associated with the Rogers-Ramanujan identities ! 
Also this coincides with the minimal model Virasoro character $\chi_{1,1}^{2,5}(q)$ up to the normalization factor. 

%modular
Under the action of the modular transformation, the normalized Rogers-Ramanujan function 
\begin{align}
\widetilde{H}(q)&=q^{11/60}H(q)
\end{align}
transforms as
\begin{align}
\widetilde{H}\left(\frac{a\tau+b\tau}{c\tau+d}\right)
&=e^{2\pi i\beta(a,b,c)/60}\widetilde{H}(\tau), 
\end{align}
where 
\begin{align}
\beta(a,b,c)&=a(3+11b+c)-3. 
\end{align}

%Rogers Ramanujan continued fraction
Accordingly, we find the elegant expression of the unflavored normalized one-point function 
in terms of the Rogers-Ramanujan continued fraction. 
It follows that 
\begin{align}
\langle \mathcal{W}_1\rangle^{SU(2)_{-5/2}}(x=1;q)
&:=\frac{\langle W_1\rangle^{SU(2)_{-5/2}}(x=1;q)}{\mathbb{II}_{\mathcal{N}}^{SU(2)_{-5/2}}(x=1;q)}
\nonumber\\
&=\frac{-q^{1/2}}{1-\frac{q}{1+\frac{q^2}{1+\frac{q^3}{1+\ddots}}}}. 
\end{align}

%proof
The unflavored result can be shown analytically as follows. 
%\textcolor{red}{
%[NOTE] 
%To prove the result with flavored fugacity, 
%we may use the following identity: 
%\begin{align}
%(q^{1/2}x;q)_{\infty}&=\frac{1}{(q;q)_{\infty}}
%\sum_{n\ge0}(-1)^n q^{n^2/2}x^n (q^{1+n};q)
%\end{align}
%}
Indeed for $x = 1$ we can use the Jacobi triple product formula \eqref{JacobiTripleProduct} twice to calculate
\begin{align}
    I_{\alpha} & = \frac{(q)_{\infty}}{2} \oint \frac{ds}{2 \pi i s} (s^{\pm 2}; q)_{\infty} (q^{1/2} s^{\pm}; q)_{\infty} s^{\alpha}
    \nonumber \\
    & = \frac{1}{2} \oint \frac{ds}{2\pi i s} \sum_{m, n \in \Zb} (-1)^{m + n} q^{m(m + 1)/2 + n^2/2} (1 - s^2)s^{2m + n + \alpha}
    \nonumber \\
    & = \frac{1}{2} \sum_{m \in \Zb} (-1)^{m + \alpha} q^{\frac{5}{2}m^2} \left( q^{(2 \alpha + 1/2)m + \alpha^2/2} - q^{(2 \alpha + 9/2)m + \alpha^2/2 + 2\alpha + 2} \right) \; .
\end{align}

The form of the result depends on $\alpha \mod 5$ so we take $\alpha = 5 \lambda + \beta$ where $\lambda \in \Zb$ and $\beta \in \{-2, -1, 0, 1, 2\}$. 
We then find the one-point of the charged Wilson line
\begin{align}
    \langle W_{\alpha} \rangle^{SU(2)_{-5/2}}  = I_{\alpha} + I_{- \alpha} & = \left\{
    \begin{array}{l}
    (-1)^{\lambda + 1} q^{\frac{5}{2}\lambda^2} \left(q^{-\lambda}G(q) + q^{-3\lambda + 1}H(q) \right)\; , \;\; \beta = -2
    \nonumber \\
    (-1)^{\lambda + 1} q^{\frac{5}{2}\lambda^2 - 2\lambda + \frac{1}{2}} H(q)\; , \;\; \beta = -1
    \nonumber \\
    (-1)^{\lambda} q^{\frac{5}{2}\lambda^2} \left(q^{-\lambda} + q^{\lambda}\right)G(q)\; , \;\; \beta = 0
    \nonumber \\
    (-1)^{\lambda + 1} q^{\frac{5}{2}\lambda^2 + 2\lambda + \frac{1}{2}} H(q)\; , \;\; \beta = 1
    \nonumber \\
    (-1)^{\lambda + 1} q^{\frac{5}{2}\lambda^2} \left(q^{\lambda}G(q) + q^{3\lambda + 1}H(q) \right)\; , \;\; \beta = 2
    \nonumber \\
    \end{array}
    \right.
\end{align}

The above gives a derivation for $\alpha = 0$ of the half-index result in the previous subsection while for $\alpha = 1$ we derive the one-point function $\langle W_1 \rangle$ above.
Some other examples are
\begin{align}
\langle W_2\rangle^{SU(2)_{-5/2}}&=-G(q)-qH(q), \\
\langle W_3\rangle^{SU(2)_{-5/2}}&=q^{3/2}G(q)+q^{1/2}H(q), \\
\langle W_4\rangle^{SU(2)_{-5/2}}&=qH(q), \\
\langle W_5\rangle^{SU(2)_{-5/2}}&=-q^{3/2}G(q)-q^{7/2}G(q), \\
\langle W_6\rangle^{SU(2)_{-5/2}}&=q^5 H(q), \\
\langle W_7\rangle^{SU(2)_{-5/2}}&=q^{7/2}G(q) + q^{13/2}H(q), \\
\langle W_8\rangle^{SU(2)_{-5/2}}&=-q^{8}G(q)-q^{5}H(q), \\
\langle W_9\rangle^{SU(2)_{-5/2}}&=-q^{13/2}H(q), \\
\langle W_{10}\rangle^{SU(2)_{-5/2}}&= q^{8}G(q) + q^{12}G(q), \\
\langle W_{11}\rangle^{SU(2)_{-5/2}}&=-q^{29/2}H(q), \\
\langle W_{12}\rangle^{SU(2)_{-5/2}}&=-q^{12}G(q)-q^{17}H(q), \\
\langle W_{13}\rangle^{SU(2)_{-5/2}}&=q^{39/2}G(q)+q^{29/2}H(q), \\
\langle W_{14}\rangle^{SU(2)_{-5/2}}&=q^{17}H(q), \\
\langle W_{15}\rangle^{SU(2)_{-5/2}}&=-q^{39/2}G(q)-q^{51/2}G(q). 
\end{align}

As for the case of $SU(2)_{-3}$ we can express the rank-$\alpha$ symmetric representation results as a sum of these results giving
\begin{align}
\langle W_{(1)}\rangle^{SU(2)_{-5/2}}&=-q^{1/2}H(q), \\
\langle W_{(2)}\rangle^{SU(2)_{-5/2}}&=-qH(q), \\
\langle W_{(3)}\rangle^{SU(2)_{-5/2}}&=q^{3/2}G(q), \\
\langle W_{(4)}\rangle^{SU(2)_{-5/2}}&=0, \\
\langle W_{(5)}\rangle^{SU(2)_{-5/2}}&=-q^{7/2}G(q), \\
\langle W_{(6)}\rangle^{SU(2)_{-5/2}}&=q^5 H(q), \\
\langle W_{(7)}\rangle^{SU(2)_{-5/2}}&= q^{13/2}H(q), \\
\langle W_{(8)}\rangle^{SU(2)_{-5/2}}&=-q^{8}G(q), \\
\langle W_{(9)}\rangle^{SU(2)_{-5/2}}&= 0, \\
\langle W_{(10)}\rangle^{SU(2)_{-5/2}}&= q^{12}G(q), \\
\langle W_{(11)}\rangle^{SU(2)_{-5/2}}&= -q^{29/2}H(q), \\
\langle W_{(12)}\rangle^{SU(2)_{-5/2}}&= -q^{17}H(q), \\
\langle W_{(13)}\rangle^{SU(2)_{-5/2}}&= q^{39/2}G(q), \\
\langle W_{(14)}\rangle^{SU(2)_{-5/2}}&= 0, \\
\langle W_{(15)}\rangle^{SU(2)_{-5/2}}&= -q^{51/2}G(q). 
\end{align}
with general results
\begin{align}
    \langle W_{(\alpha)} \rangle^{SU(2)_{-5/2}}  & = \left\{
    \begin{array}{l}
    (-1)^{\lambda + 1} q^{\frac{5}{2}\lambda^2 - \lambda}G(q)\; , \;\; \beta = -2
    \nonumber \\
    0\; , \;\; \beta = -1
    \nonumber \\
    (-1)^{\lambda} q^{\frac{5}{2}\lambda^2 + \lambda} G(q)\; , \;\; \beta = 0
    \nonumber \\
    (-1)^{\lambda + 1} q^{\frac{5}{2}\lambda^2 + 2\lambda + \frac{1}{2}} H(q)\; , \;\; \beta = 1
    \nonumber \\
    (-1)^{\lambda + 1} q^{\frac{5}{2}\lambda^2 + 3\lambda + 1}H(q)\; , \;\; \beta = 2
    \nonumber \\
    \end{array}
    \right.
\end{align}

%%%%%%%%%%%%%%%%%%%%%%%%%%%%%%%%%%
%%%%%%%%%%%%%%%%%%%%%%%%%%%%%%%%%%
\subsection{$k = -3$}
\label{sec_su2_km3}
%%%%%%%%%%%%%%%%%%%%%%%%%%%%%%%%%%
%%%%%%%%%%%%%%%%%%%%%%%%%%%%%%%%%%
Taking $N_f = 0$ and $M = 1$ (or alternatively $M = 0$ with $2$ 3d chirals with Dirichlet boundary conditions but with the axial fugacity set to $1$ in the half-index below) gives $k = -3$. The half-index reads
\begin{align}
\label{h_su2km3}
\mathbb{II}_{\mathcal{N}}^{SU(2)_{-3}}(x;q)
&=\frac{(q;q)_{\infty}}{2} 
\oint \frac{ds}{2 \pi i s} (s^{\pm 2}; q)_{\infty} 
(q^{\frac12}sx^{\pm};q)_{\infty}
(q^{\frac12}s^{-1}x^{\mp};q)_{\infty}. 
\end{align}
It can be expanded as
\begin{align}
&1+(1+x^2+x^{-2})q+(2+x^{2}+x^{-2})q^2+(3+2x^2+2x^{-2})q^3
\nonumber\\
&+(5+x^4+3x^2+3x^{-2}+x^{-4})q^4+(7+x^4+5x^2+5x^{-2}+x^{-4})q^5+\cdots
\end{align}
%\textcolor{red}{[NOTE] Physical interpretation}
Here the detailed interpretation depends on whether we consider two fundamental 3d chirals or a single 2d Fermi but as in previous cases we have a fairly straightforward counting of gauge invariants built from the two fundamental chirals, taking into account symmetry properties.

We find that the half-index (\ref{h_su2km3}) is given by
\begin{align}
&\frac{1}{(q;q)_{\infty}}
\sum_{n\in \mathbb{Z}}
q^{n^2}x^{2n}
=
e^{\frac{\pi i\tau}{12}}
\frac{\vartheta_3(2z;2\tau)}{\eta(\tau)}, 
\end{align}
where $\vartheta_3(z;\tau)$ is the Jacobi theta function (\ref{theta3}) 
and $\eta(\tau)$ is the Dedekind eta function (\ref{eta}). 
This is the vacuum character of the $U(1)_{2}$ WZW model
consistent with the duality of boundary conditions proposed in \cite{Dimofte:2017tpi}
\begin{align}
&\textrm{$SU(2)_{-3}$ pure CS with Neumann b.c. $+$ fund. Fermi}
\nonumber\\
&\leftrightarrow 
\textrm{$U(1)_{2}$ pure CS with Dirichlet b.c.}
\end{align}

When the flavored fugacity $x$ is turned off, we get
\begin{align}
\label{fcn_A}
A(q)&:=\frac{1}{(q;q)_{\infty}}\left(
1+2\sum_{n=1}^{\infty} q^{n^2}\right)
\nonumber\\
&=\prod_{n=1}^{\infty}
\frac{(1+q^{2n-1}) (1-q^{2n})}
{(1-q^{2n-1}) (1+q^{2n})(1-q^{n})}
\nonumber\\
&=\frac{\varphi(q)}{f(-q)}, 
\end{align}
where $\varphi(q)$ and $f(-q)$ are Ramanujan's theta functions (\ref{r_theta_phi}) and (\ref{r_theta_f}). 

%%%%%%%%%%%%%%%%%%%%%%%%%%%%%%%%%%
\subsubsection{One-point functions}
\label{sec_1pt_su2km3}
%%%%%%%%%%%%%%%%%%%%%%%%%%%%%%%%%%
The one-point function of the Wilson line operator in the fundamental representation is evaluated as 
\begin{align}
\label{w1_su2km3}
\langle W_1\rangle^{SU(2)_{-3}}(x;q)
&=\frac{(q;q)_{\infty}}{2}
 \oint \frac{ds}{2 \pi i s} (s^{\pm 2}; q)_{\infty} 
(q^{\frac12}sx^{\pm};q)_{\infty}
(q^{\frac12}s^{-1}x^{\mp};q)_{\infty}
(s+s^{-1}). 
\end{align}
It has an expansion
\begin{align}
&-(x+x^{-1})q^{1/2}-(x+x^{-1})q^{3/2}-(x^3+2x+2x^{-1}+x^{-3})q^{5/2}
\nonumber\\
&-(x^3+3x+3x^{-1}+x^{-3})q^{7/2}-(2x^3+5x+5x^{-1}+2x^{-3})q^{9/2}-\cdots.
\end{align}
The one-point function should count the BPS boundary local operators at the insertion point of the fundamental Wilson line. 
We see that only fermionic local operators can attach with the fundamental Wilson line since the Wilson line itself is in the spin-$\frac{1}{2}$ representation of $SU(2)$ so to make a gauge singlet the additional boundary local operator must also be in an odd half-integer spin representation. The only way to form such representations is from an odd number of the 3d chiral multiplets. Since they have Dirichlet boundary conditions the bosonic component is fixed at the boundary so the degrees of freedom lie in the fermionic component hence the local operator formed in this way must be fermionic.
%\textcolor{red}{[NOTE] Physical interpretation is actually more complicated as there are gauginos and there are partial cancellations in the half-index.}

The one-point function matches 
\begin{align}
\langle W_1\rangle^{SU(2)_{-3}}(x;q)
&=-\frac{1}{(q;q)_{\infty}}
\sum_{n\in \mathbb{Z}}
q^{n^2+n+\frac12}x^{2n+1}. 
%&=-\frac{q^{\frac12}}{(q;q)_{\infty}}
%\sum_{n=0}^{\infty} q^{n^2 + n}(x^{2n+1}+x^{-2n-1}). 
\end{align}

The half-index (\ref{h_su2km3}) and the one-point function (\ref{w1_su2km3}) satisfies the $q$-difference equation
\begin{align}
\label{qdiff_su2k3}
\mathbb{II}_{\mathcal{N}}^{SU(2)_{-3}}(q^{1/2}x;q)&=
-q^{-1/2}x^{-1}\langle W_1\rangle^{SU(2)_{-3}}(x;q)
\end{align}
which takes the same form as (\ref{qdiff_su2k5/2}).

In the unflavored limit $x\rightarrow 1$, the one-point function reduces to 
\begin{align}
-2q^{1/2} B(q), 
\end{align}
where 
\begin{align}
\label{fcn_B}
B(q)&:=\frac{1}{(q;q)_{\infty}}\sum_{n=0}^{\infty}q^{n^2+n}
\nonumber\\
&=\prod_{n=1}^{\infty} \frac{1-q^{4n}}{(1-q^{4n-2}) (1-q^{n})}
\nonumber\\
&=\frac{\psi(q^2)}{f(-q)}. 
\end{align}
Here $\psi(q)$ and $f(-q)$ are Ramanujan's theta functions (\ref{r_theta_psi}) and (\ref{r_theta_f}). 

%proof
We give an analytic proof in the following. 
Indeed, using the Jacobi triple product formula three times we have
\begin{align}
V_{\alpha} & \equiv
\frac{1}{2} (q;q)_{\infty} \oint \frac{ds}{2 \pi i s} (s^{\pm 2}; q)_{\infty} 
(q^{\frac12}sx^{\pm};q)_{\infty}
(q^{\frac12}s^{-1}x^{\mp};q)_{\infty}
s^{\alpha}
\nonumber \\
& = \frac{1}{2(q;q)_{\infty}^2} \oint \frac{ds}{2 \pi i s} \sum_{m, n, l \in \Zb} (-1)^{l + m + n} q^{l(l+1)/2 + m^2/2 + n^2/2} (1 - s^2) s^{2l + m + n + \alpha} x^{m - n}
\nonumber \\
& =I_{\alpha}(x; q) - I_{\alpha + 2}(x; q)
\end{align}
where
\begin{align}
I_{\alpha}(x; q) & = \frac{1}{2(q;q)_{\infty}^2} \sum_{m, l \in \Zb} (-1)^{l + \alpha} q^{P_{\alpha}} x^{2(m + l)} \; , \\
P_{\alpha} & = \frac{3}{2}l^2 + \frac{1}{2}l + (m + l)^2 + \alpha(m + l) + \alpha l + \frac{1}{2}\alpha^2 \; .
\end{align}
Now we define $M = m + l$ and, $L = l + \alpha/3$ for $\alpha \equiv 0 \mod 3$, while $L = -l - (\alpha + 1)/3$ for $\alpha \equiv -1 \mod 3$. In both these cases we then find
\begin{align}
I_{\alpha}(x; q) & = \hat{S}_{\alpha} \frac{1}{2(q;q)_{\infty}^2} \sum_{M, L \in \Zb} (-1)^{L} q^{\frac{3}{2}L^2 + \frac{1}{2}L + M^2 + \alpha M + \frac{\alpha^2}{3} - \frac{\alpha}{6}} x^{2M}
\nonumber \\
 & = \hat{S}_{\alpha} \frac{1}{2(q;q)_{\infty}} \sum_{M \in \Zb} q^{M^2 + \alpha M + \frac{\alpha^2}{3} - \frac{\alpha}{6}} x^{2M}
\end{align}
where $\hat{S}_{\alpha} = 1$ for $\alpha \equiv 0 \mod 3$ and $\hat{S}_{\alpha} = -1$ for $\alpha \equiv -1 \mod 3$.

Now note that
\begin{align}
    M^2 + \alpha M + \frac{\alpha^2}{3} - \frac{\alpha}{6}
     & = (M + \frac{\alpha}{2})^2 + \frac{\alpha^2}{12} - \frac{\alpha}{6} \\
     & = (M + \frac{\alpha - 1}{2})^2 + (M + \frac{\alpha - 1}{2}) + \frac{\alpha^2}{12} - \frac{\alpha}{6} + \frac{1}{4}, 
\end{align}
so for $\alpha \equiv 0 \mod 2$ we have
\begin{align}
I_{\alpha}(x; q) & = \hat{S}_{\alpha} x^{-\alpha} \frac{1}{2(q;q)_{\infty}} \sum_{M \in \Zb} q^{M^2 + \frac{\alpha^2}{12} - \frac{\alpha}{6}} x^{2M}
\nonumber \\
 & \equiv \frac{1}{2} \hat{S}_{\alpha} x^{-\alpha} q^{\frac{\alpha^2}{12} - \frac{\alpha}{6}} A(x; q), 
\end{align}
while for $\alpha \equiv 1 \mod 2$ we have
\begin{align}
I_{\alpha}(x; q) & = \hat{S}_{\alpha} x^{1 - \alpha} \frac{1}{2(q;q)_{\infty}} \sum_{M \in \Zb} q^{M^2 + M + \frac{\alpha^2}{12} - \frac{\alpha}{6} + \frac{1}{4}} x^{2M}
 \nonumber \\
 & \equiv \hat{S}_{\alpha} x^{1 - \alpha} q^{\frac{\alpha^2}{12} - \frac{\alpha}{6} + \frac{1}{4}} B(x; q). 
\end{align}

Similarly, for $\alpha \equiv 1 \mod 3$ we can write $\alpha = 3 \beta + 1$ where $\beta \in \Zb$ and we have
\begin{align}
    P_{\alpha} & = \frac{3}{2}(l + \beta)(l + \beta + 1) + (m + l)^2 + (3\beta + 1)(m + l) + 3\beta^2 + \frac{3}{2}\beta + \frac{1}{2}
\nonumber \\
    & = \frac{3}{2}(-l - \beta - 1)((-l - \beta - 1) + 1) + (m + l)^2 + (3\beta + 1)(m + l) + 3\beta^2 + \frac{3}{2}\beta + \frac{1}{2}, 
\end{align}
so due to the factor $(-1)^{l + \alpha} = (-1)^{\alpha + \beta}(-1)^{l + \beta}  = -(-1)^{\alpha + \beta}(-1)^{-l - \beta - 1}$
we see that $I_{\alpha} = 0$ for $\alpha \equiv 1 \mod 3$.

Now we can calculate the one-point function of the charge-$\alpha$ Wilson line
\begin{align}
    \langle W_{\alpha} \rangle & = V_{\alpha} + V_{- \alpha}
     = I_{\alpha} - I_{\alpha + 2} + I_{- \alpha} - I_{- \alpha + 2}
\end{align}
From above we see that the details depend on $\alpha \mod 6$. Specifically, we find
\begin{align}
    \langle W_{\alpha} \rangle^{SU(2)_{-3}} & = \left\{ \begin{array}{ll}
    \frac{1}{2}(x^{-\alpha} + x^{\alpha - 2} + q^{\alpha/3} x^{\alpha} + q^{\alpha/3} x^{-\alpha - 2})q^{\alpha(\alpha - 2)/12} A(x; q) & \mathrm{for} \; \alpha = 0 \mod 6
    \nonumber \\
    -(x^{-\alpha - 1} + x^{\alpha + 1})q^{\alpha(\alpha + 2)/12 + 1/4} B(x; q) & \mathrm{for} \; \alpha = 1 \mod 6
    \nonumber \\
    -\frac{1}{2}(x^{-\alpha} + x^{\alpha - 2})q^{\alpha(\alpha - 2)/12} A(x; q) & \mathrm{for} \; \alpha = 2 \mod 6
    \nonumber \\
    (x^{1 - \alpha} + x^{\alpha - 1} + q^{\alpha/3} x^{\alpha + 1} + q^{\alpha/3} x^{-\alpha - 1})q^{\alpha(\alpha - 2)/12 + 1/4} B(x; q) & \mathrm{for} \; \alpha = 3 \mod 6
    \nonumber \\
    -\frac{1}{2}(x^{-\alpha - 2} + x^{\alpha})q^{\alpha(\alpha + 2)/12} A(x; q) & \mathrm{for} \; \alpha = 4 \mod 6
    \nonumber \\
    -(x^{1 - \alpha} + x^{\alpha - 1})q^{\alpha(\alpha - 2)/12 + 1/4} B(x; q) & \mathrm{for} \; \alpha = 5 \mod 6
    \nonumber \\
    \end{array} \right.
\end{align}

It follows that for $x = 1$ the correlators are expressible in terms of the functions $A(q) = A(1; q)$, (\ref{fcn_A}) and $B(q) = B(1; q)$, (\ref{fcn_B}). 
We have
\begin{align}
    \langle W_{\alpha} \rangle^{SU(2)_{-3}} & = \left\{ \begin{array}{ll}
    (1 + q^{\alpha/3})q^{\alpha(\alpha - 2)/12} A(q) & \mathrm{for} \; \alpha = 0 \mod 6
    \nonumber \\
    -2q^{\alpha(\alpha + 2)/12 + 1/4} B(q) & \mathrm{for} \; \alpha = 1 \mod 6
    \nonumber \\
    -q^{\alpha(\alpha - 2)/12} A(q) & \mathrm{for} \; \alpha = 2 \mod 6
    \nonumber \\
    2(1 + q^{\alpha/3})q^{\alpha(\alpha - 2)/12 + 1/4} B(q) & \mathrm{for} \; \alpha = 3 \mod 6
    \nonumber \\
    -q^{\alpha(\alpha + 2)/12} A(q) & \mathrm{for} \; \alpha = 4 \mod 6
    \nonumber \\
    -2q^{\alpha(\alpha - 2)/12 + 1/4} B(q) & \mathrm{for} \; \alpha = 5 \mod 6
    \nonumber \\
    \end{array} \right.
\end{align}
For example
\begin{align}
\langle W_2\rangle^{SU(2)_{-3}}&=-A(q), \\
\langle W_3\rangle^{SU(2)_{-3}}&=2q^{1/2}B(q)+2q^{3/2}B(q), \\
\langle W_4\rangle^{SU(2)_{-3}}&=-q^2A(q), \\
\langle W_5\rangle^{SU(2)_{-3}}&=-2q^{3/2}B(q), \\
\langle W_6\rangle^{SU(2)_{-3}}&=q^{2}A(q)+q^4A(q), \\
\langle W_7\rangle^{SU(2)_{-3}}&=-2q^{\frac{11}{2}}B(q), \\
\langle W_8\rangle^{SU(2)_{-3}}&=-q^4A(q), \\
\langle W_9\rangle^{SU(2)_{-3}}&=2q^{11/2}B(q)+2q^{17/2}B(q), \\
\langle W_{10}\rangle^{SU(2)_{-3}}&=-q^{10}A(q), \\
\langle W_{11}\rangle^{SU(2)_{-3}}&=-2q^{17/2}B(q), \\
\langle W_{12}\rangle^{SU(2)_{-3}}&=q^{10}A(q)+q^{14}A(q), \\
\langle W_{13}\rangle^{SU(2)_{-3}}&=-2q^{\frac{33}{2}}B(q), \\
\langle W_{14}\rangle^{SU(2)_{-3}}&=-q^{14}A(q), \\
\langle W_{15}\rangle^{SU(2)_{-3}}&=2q^{33/2}B(q)+2q^{43/2}B(q). 
\end{align}

For $SU(2)$ the Wilson line half-index for the rank-$n$ symmetric representation for $n$ odd (even) is given by the sum of the charge-odd (charge-even) Wilson line half-indices, where for charge-$0$ we instead use the half-index without a Wilson line.
For example
\begin{align}
\langle W_{(1)}\rangle^{SU(2)_{-3}}&=-2q^{1/2}B(q), \\
\langle W_{(2)}\rangle^{SU(2)_{-3}}&=0, \\
\langle W_{(3)}\rangle^{SU(2)_{-3}}&=2q^{3/2}B(q), \\
\langle W_{(4)}\rangle^{SU(2)_{-3}}&=-q^2A(q), \\
\langle W_{(5)}\rangle^{SU(2)_{-3}}&=0, \\
\langle W_{(6)}\rangle^{SU(2)_{-3}}&=q^4A(q), \\
\langle W_{(7)}\rangle^{SU(2)_{-3}}&=-2q^{\frac{11}{2}}B(q), \\
\langle W_{(8)}\rangle^{SU(2)_{-3}}&=0, \\
\langle W_{(9)}\rangle^{SU(2)_{-3}}&=2q^{17/2}B(q), \\
\langle W_{10)}\rangle^{SU(2)_{-3}}&=-q^{10}A(q), \\
\langle W_{(11)}\rangle^{SU(2)_{-3}}&=0, \\
\langle W_{(12)}\rangle^{SU(2)_{-3}}&=q^{14}A(q), \\
\langle W_{(13)}\rangle^{SU(2)_{-3}}&=-2q^{\frac{33}{2}}B(q), \\
\langle W_{(14)}\rangle^{SU(2)_{-3}}&=0, \\
\langle W_{(15)}\rangle^{SU(2)_{-3}}&=2q^{43/2}B(q). 
\end{align}
with the general result
\begin{align}
    \langle W_{(\alpha)} \rangle^{SU(2)_{-3}} & = \left\{ \begin{array}{ll}
    q^{\alpha(\alpha - 2)/12} A(q) & \mathrm{for} \; \alpha = 0 \mod 6
    \nonumber \\
    -2q^{\alpha(\alpha + 2)/12 + 1/4} B(q) & \mathrm{for} \; \alpha = 1 \mod 6
    \nonumber \\
    0 & \mathrm{for} \; \alpha = 2 \mod 6
    \nonumber \\
    2q^{\alpha(\alpha + 2)/12 + 1/4} B(q) & \mathrm{for} \; \alpha = 3 \mod 6
    \nonumber \\
    -q^{\alpha(\alpha + 2)/12} A(q) & \mathrm{for} \; \alpha = 4 \mod 6
    \nonumber \\
    0 & \mathrm{for} \; \alpha = 5 \mod 6
    \nonumber \\
    \end{array} \right.
\end{align}

%%%%%%%%%%%%%%%%%%%%%%%%%%%%%%%%%%
%%%%%%%%%%%%%%%%%%%%%%%%%%%%%%%%%%
\subsection{$k= -3/2 - M$}
\label{sec_su2_km32M}
%%%%%%%%%%%%%%%%%%%%%%%%%%%%%%%%%%
%%%%%%%%%%%%%%%%%%%%%%%%%%%%%%%%%%
With $N_f = 1$ and $M$ Fermis, we expect the Neumann half-indices are equal to half-indices of dual $U(M)_{3/2 + M, 3/2}$ theories with Dirichlet boundary conditions for the gauge multiplet and a 3d fundamental chiral. We have already described the case of $M = 1$ (although with a specialisation of the global $U(1)$ fugacities equivalent to instead taking a single fundamental chiral with Dirichlet boundary conditions and no Fermis) and here we list some further examples for $M = 2$ and $M = 3$. However, the one-point functions of Wilson lines operators are more complicated and we don't have any general conjectures or $q$-difference equations relating these one-point functions to the half-indices in the absence of Wilson lines. Most likely there should be some useful and interesting $q$-difference equations but we leave that to future works.

%%%%%%%%%%%%%%%%%%%%%%%%%%%%%%%%%%
\subsubsection{$k= - 7/2$}
\label{sec_su2_km72}
%%%%%%%%%%%%%%%%%%%%%%%%%%%%%%%%%%
While there are multiple choices of the matter content to get the Chern-Simons level $k=-7/2$, 
let us consider the case with two fundamental Fermi multiplets as well as a single chiral multiplet satisfying the Neumann boundary condition. 
The half-index is given by
\begin{align}
\label{h_su2km72}
\mathbb{II}_{\mathcal{N}}^{SU(2)_{-7/2}}(x_{\alpha};q)
&=\frac{(q;q)_{\infty}}{2}
\oint \frac{ds}{2\pi is}
(s^{\pm2};q)_{\infty}
\frac{\prod_{\alpha=1}^2 (q^{\frac12}s^{\pm}x_{\alpha}^{\pm};q) (q^{\frac12}s^{\mp}x_{\alpha}^{\pm};q)}
{(q^{\frac12}s^{\pm}x_3;q)_{\infty}}
\end{align}
where $x_1,x_2$ are the fugacities for the flavor symmetry rotating the two Fermi multiplet 
and $x_3$ is the fugacity for the chiral multiplet. 
It can be expanded as 
\begin{align}
\mathbb{II}_{\mathcal{N}}^{SU(2)_{-7/2}}
&=1+\Bigl(
2+x_1^2+x_2^2+x_1^{-2}+x_2^{-2}
+x_1x_2+x_1^{-1}x_2^{-1}+x_1^{-1}x_2+x_1x_2^{-1}
\nonumber\\
&-x_1x_3-x_2x_3-x_3x_1^{-1}-x_3x_2^{-1}
\Bigr)q+\cdots
\end{align}
%\textcolor{red}{
%[NOTE]: Physical interpretation, making comments on the fermionic operators. 
%It will be distinguished from the case with integer levels.
%}
Here gauge invariants can be formed by pairing two Fermi multiplet contributions or one Fermi multiplet with the 3d chiral. The Fermis contribute fermionic terms while the Neumann chiral contributes a bosonic contribution.

We find that the half-index (\ref{h_su2km72}) coincides with
\begin{align}
\label{h_dualsu2km72}
\frac{1}{(q;q)_{\infty}^2}
\sum_{m_1,m_2\in \mathbb{Z}}
\frac{q^{m_1^2+m_2^2}x_1^{2m_1}x_2^{2m_2} (q^{1-m_1}x_1^{-1}x_3;q)_{\infty} (q^{1-m_2}x_2^{-1}x_3;q)_{\infty}}
{(q^{1\pm m_1\mp m_2}x_1^{\pm}x_2^{\mp};q)_{\infty}}. 
\end{align}
The series expression (\ref{h_dualsu2km72}) indicates that, as expected, the dual description is given by 
the $U(2)_{7/2, 3/2}$ Chern-Simons theory with a fundamental 3d chiral with Dirichlet boundary conditions.

%%%%%%%%%%%%%%%%%%%%%%%%%%%%%%%%%%
%%%%%%%%%%%%%%%%%%%%%%%%%%%%%%%%%%
\subsubsection{$k= - 9/2$}
\label{sec_su2_km92}
%%%%%%%%%%%%%%%%%%%%%%%%%%%%%%%%%%
%%%%%%%%%%%%%%%%%%%%%%%%%%%%%%%%%%
For the Chern-Simons level $k=-9/2$, 
one can take three fundamental Fermi multiplets along with a single chiral multiplet obeying the Neumann boundary condition. 
The half-index is given by
\begin{align}
\label{h_su2km92}
\mathbb{II}_{\mathcal{N}}^{SU(2)_{-9/2}}(x_{\alpha};q)
&=\frac{(q;q)_{\infty}}{2}
\oint \frac{ds}{2\pi is}
(s^{\pm2};q)_{\infty}
\frac{\prod_{\alpha=1}^3 (q^{\frac12}s^{\pm}x_{\alpha}^{\pm};q) (q^{\frac12}s^{\mp}x_{\alpha}^{\pm};q)}
{(q^{\frac12}s^{\pm}x_4;q)_{\infty}}. 
\end{align}
We find that the half-index (\ref{h_su2km92}) matches with
\begin{align}
\label{h_dualsu2km92}
\frac{1}{(q;q)_{\infty}^3}\sum_{m_1,m_2,m_3\in \mathbb{Z}}
\frac{q^{m_1^2+m_2^2+m_3^2} x_1^{2m_1}x_2^{2m_2}x_3^{2m_3}
\prod_{\alpha=1}^3 (q^{1-m_{\alpha}}x_{\alpha}^{-1}x_4;q)_{\infty}
}
{\prod_{\alpha<\beta}^{3}(q^{1\pm m_{\alpha}\mp m_{\beta}}x_{\alpha}^{\pm} x_{\beta}^{\mp};q)_{\infty}}. 
\end{align}
The series expression (\ref{h_dualsu2km92}) indicates that the dual description is given, as expected, by 
the $U(3)_{9/2, 3/2}$ Chern-Simons theory with a fundamental chiral having Dirichlet boundary conditions.

%%%%%%%%%%%%%%%%%%%%%%%%%%%%%%%%%%
%%%%%%%%%%%%%%%%%%%%%%%%%%%%%%%%%%
\subsection{$k = -2 - M$}
\label{sec_su2_km2M}
%%%%%%%%%%%%%%%%%%%%%%%%%%%%%%%%%%
%%%%%%%%%%%%%%%%%%%%%%%%%%%%%%%%%%
We expect the Neumann half-indices are equal to half-indices of dual $U(M)_{2 + M, 2}$ theories with Dirichlet boundary conditions for the gauge multiplet and a 3d fundamental chiral. We have already described the case of $M = 1$ and here we list some further examples of dual half-indices as well as a discussion of the cases with a line operator. These cases
match the duality of boundary conditions proposed in \cite{Dimofte:2017tpi}
\begin{align}
&\textrm{$SU(2)_{-2 - M}$ pure CS with Neumann b.c. $+$ $M$ fund. Fermis}
\nonumber\\
&\leftrightarrow 
\textrm{$U(M)_{2 + M,2}$ pure CS with Dirichlet b.c.}
\end{align}

%%%%%%%%%%%%%%%%%%%%%%%%%%%%%%%%%%
\subsubsection{$k=-4$}
\label{sec_su2_km4}
%%%%%%%%%%%%%%%%%%%%%%%%%%%%%%%%%%
When $N_f = N_a = 0$ and $M=2$, we have $k=-4$. 
The half-index is given by
\begin{align}
\label{h_su2km4}
\mathbb{II}_{\mathcal{N}}^{SU(2)_{-4}}(x_{\alpha};q)
&=\frac{(q;q)_{\infty}}{2}
\oint \frac{ds}{2\pi is}
(s^{\pm2};q)_{\infty}
\prod_{\alpha=1}^{2}
(q^{\frac12}s^{\pm}x_{\alpha}^{\pm};q)_{\infty}
(q^{\frac12}s^{\pm}x_{\alpha}^{\mp};q)_{\infty}. 
\end{align}
We find that the half-index (\ref{h_su2km4}) matches 
\begin{align}
\label{h_wzwU(2)_2}
\frac{1}{(q;q)_{\infty}^2}\sum_{m_1,m_2\in \mathbb{Z}}
\frac{q^{m_1^2+m_2^2} x_1^{2m_1}x_2^{2m_2}}
{(q^{1\pm m_1\mp m_2}x_1^{\pm} x_2^{\mp};q)_{\infty}}. 
\end{align}
The expression (\ref{h_wzwU(2)_2}) can be interpreted as 
the vacuum character of the $U(2)_{2}$ WZW model or the pure $U(2)_{4, 2}$ Chern-Simons theory.

%%%%%%%%%%%%%%%%%%%%%%%%%%%%%%%%%%
\subsubsection{$k=-4$ One-point function}
\label{sec_1pt_su2km4}
%%%%%%%%%%%%%%%%%%%%%%%%%%%%%%%%%%

We find that the one-point function of the Wilson line in the fundamental representation coincides with
\begin{align}
\label{w1_su2km4}
\langle W_1\rangle^{SU(2)_{-4}}(x_{\alpha};q)
&=-\frac{q^{1/2}}{(q;q)_{\infty}^2}
\sum_{m_1,m_2\in \mathbb{Z}}
\frac{q^{m_1^2+m_2^2} x_1^{2m_1} x_2^{2m_2} (q^{m_1}x_1+q^{m_2}x_2)}
{(q^{1\pm m_1\mp m_2}x_1^{\pm} x_2^{\mp};q)_{\infty}}.
\end{align}
Accordingly, we obtain the $q$-difference equation satisfied by the half-index (\ref{h_su2km4}) and (\ref{w1_su2km4})
\begin{align}
&
\langle W_1\rangle^{SU(2)_{-4}}(x_1,x_2;q)
\nonumber\\
&=-q^{1/2}x_1 \mathbb{II}_{\mathcal{N}}^{SU(2)_{-4}}(q^{1/2}x_1,x_2;q)
-q^{1/2}x_2 \mathbb{II}_{\mathcal{N}}^{SU(2)_{-4}}(x_1,q^{1/2}x_2;q). 
\end{align}

%%%%%%%%%%%%%%%%%%%%%%%%%%%%%%%%%%
\subsubsection{$k=-5$}
\label{sec_su2_km5}
%%%%%%%%%%%%%%%%%%%%%%%%%%%%%%%%%%
For $N_f = N_A = 3$ and $M=3$, the Chern-Simons level is $k=-5$. 
The half-index is
\begin{align}
\label{h_su2km5}
\mathbb{II}_{\mathcal{N}}^{SU(2)_{-5}}(x_{\alpha};q)
&=\frac{(q;q)_{\infty}}{2}
\oint \frac{ds}{2\pi is}
(s^{\pm2};q)_{\infty}
\prod_{\alpha=1}^{3}
(q^{\frac12}s^{\pm}x_{\alpha}^{\pm};q)_{\infty}
(q^{\frac12}s^{\pm}x_{\alpha}^{\mp};q)_{\infty}. 
\end{align}
We have confirmed that it agrees with
\begin{align}
\frac{1}{(q;q)_{\infty}^3}\sum_{m_1,m_2,m_3\in \mathbb{Z}}
\frac{q^{m_1^2+m_2^2+m_3^2} x_1^{2m_1}x_2^{2m_2}x_3^{2m_3}
}
{\prod_{\alpha<\beta}^{3}(q^{1\pm m_{\alpha}\mp m_{\beta}}x_{\alpha}^{\pm} x_{\beta}^{\mp};q)_{\infty}}, 
\end{align}
which is identified with the vaccum character of the $U(3)_{2}$ WZW model
or the pure $U(3)_{5, 3}$ Chern-Simons theory.

%%%%%%%%%%%%%%%%%%%%%%%%%%%%%%%%%%
\subsubsection{$k=-5$ One-point function}
\label{sec_1pt_su2km5}
%%%%%%%%%%%%%%%%%%%%%%%%%%%%%%%%%%

We find that the one-point function of the fundamental Wilson line is given by
\begin{align}
\label{w1_su2km5}
\langle W_1\rangle^{SU(2)_{-5}}(x_{\alpha};q)
&=\frac{-q^{1/2}}{(q;q)_{\infty}^3}\sum_{m_1,m_2,m_3\in \mathbb{Z}}
\frac{q^{m_1^2+m_2^2+m_3^2} x_1^{2m_1}x_2^{2m_2}x_3^{2m_3}
(q^{m_1}x_1+q^{m_2}x_2+q^{m_3}x_3)
}
{\prod_{\alpha<\beta}^{3}(q^{1\pm m_{\alpha}\mp m_{\beta}}x_{\alpha}^{\pm} x_{\beta}^{\mp};q)_{\infty}}. 
\end{align}
The half-index (\ref{h_su2km5}) and the one-point function (\ref{w1_su2km5}) obey the $q$-difference equation
\begin{align}
&
\langle W_1\rangle^{SU(2)_{-5}}(x_1,x_2,x_3;q)
\nonumber\\
&=-q^{1/2}x_1 \mathbb{II}_{\mathcal{N}}^{SU(2)_{-5}}(q^{1/2}x_1,x_2,x_3;q)
-q^{1/2}x_2 \mathbb{II}_{\mathcal{N}}^{SU(2)_{-5}}(x_1,q^{1/2}x_2,x_3;q)
\nonumber\\
&-q^{1/2}x_3 \mathbb{II}_{\mathcal{N}}^{SU(2)_{-5}}(x_1,x_2,q^{1/2}x_3;q). 
\end{align}

%%%%%%%%%%%%%%%%%%%%%%%%%%%%%%%%%%
%%%%%%%%%%%%%%%%%%%%%%%%%%%%%%%%%%
\section{$SU(3)$}
\label{sec_su3}
%%%%%%%%%%%%%%%%%%%%%%%%%%%%%%%%%%
%%%%%%%%%%%%%%%%%%%%%%%%%%%%%%%%%%
Let us study the line defect half-indices for the 3d $\mathcal{N}=2$ $SU(3)$ Chern-Simons theory with Wilson lines and Neumann boundary conditions for the vector multiplet. 
The Chern-Simons level is fixed by
\begin{align}
\label{su3_anom_CS}
    k & = -\left( 3 + \frac{N_f+N_a}{2} + M \right) \; .
\end{align}

For gauge group $SU(3)$ we have two independent gauge fugacities $s_1$ and $s_2$ with $s_3 = s_1^{-1} s_2^{-1}$ in the computation of the half-indices and the correlation functions. 

%%%%%%%%%%%%%%%%%%%%%%%%%%%%%%%%%%
\subsection{$k=-3$}
\label{sec_su3_km3}
%%%%%%%%%%%%%%%%%%%%%%%%%%%%%%%%%%
For $N_f=0$ and $M=0$, we have the 3d $\mathcal{N}=2$ $SU(3)$ pure Chern-Simons theory with level $k=-3$. 
Since there is no matter field, the half-index is trivial. 

%%%%%%%%%%%%%%%%%%%%%%%%%%%%%%%%%%
\subsubsection{One-point function}
\label{sec_1pt_su3}
%%%%%%%%%%%%%%%%%%%%%%%%%%%%%%%%%%
We find that the one-point function of the Wilson line $W_{(k)}$ transforming in the rank-$k$ symmetric representation is given by
\begin{align}
\label{su3_m3_sym}
\langle W_{(k)}\rangle^{SU(3)_{-3}}
=
&\begin{cases}
q^{\frac{k(k+3)}{9}}&\textrm{for $k\equiv 0 \mod 3$}\cr
0&\textrm{otherwise}\cr
\end{cases}. 
\end{align}
As for the case of $SU(2)_{-2}$, this can be understood by noting that the half-index in the presence of a Wilson line in some representation counts the BPS operators which can form a gauge singlet in combination with the Wilson line representation.

Again, in this case the only boundary BPS operators are the gaugino $\lambda_{-}$ and its derivatives $D_z^n \lambda_{-}$ which are in the adjoint representation of $SU(3)$ and contribute fugacity $-q^{n+1}$ in addition to gauge fugacities. To form a gauge singlet together with the Wilson line rank-$k$ symmetric representation we need to combine these operators in a way which results in precisely $k$ anti-fundamental indices. Now, the adjoint representation has one fundamental and one anti-fundamental index but for $SU(3)$ a fundamental index is equivalent to two anti-fundamental indices. However, these two anti-fundamental indices are antisymmetrized so cannot give a non-vanishing contribution when contracted with the symmetrized Wilson line indices. Instead we need to take two adjoint operators and replace the pair of fundamental indices with a single anti-fundamental index, resulting in an operator with three anti-fundamental indices, e.g. $(\lambda_{-})^{\alpha_1}_{\beta_1} (\lambda_{-})^{\alpha_2}_{\beta_2} \epsilon_{\alpha_1 \alpha_2 \beta_3}$. We must take $k/3$ such operators to combine with the rank-$k$ symmetric representation Wilson line operator.

Since the gauginos and their derivatives are fermionic there are constraints on the number of gauginos with a specific number of derivatives in order that the contraction with the $k$ symmetrized indices from the Wilson line does not result in a vanishing operator. It turns out that the minimal choice is for each pairing to be formed from gauginos with the same number, $n$, of derivatives and for no two pairs to have the same value of $n$. For example, consider $(\lambda_{-})^{\alpha_1}_{\beta_1} (\lambda_{-})^{\alpha_2}_{\beta_2} \epsilon_{\alpha_1 \alpha_2 \beta_3} (\lambda_{-})^{\alpha_4}_{\beta_4} (D_z \lambda_{-})^{\alpha_5}_{\beta_5} \epsilon_{\alpha_4 \alpha_5 \beta_6}$. Symmetrizing over $\beta_1$, $\beta_2$ and $\beta_4$ shows that the upper indices $\alpha_1$, $\alpha_2$ and $\alpha_4$ are antisymmetrized, and hence $(\lambda_{-})^{\alpha_1}_{( \beta_1} (\lambda_{-})^{\alpha_2}_{\beta_2} (\lambda_{-})^{\alpha_4}_{\beta_4 )} = \epsilon^{\alpha_1 \alpha_2 \alpha_4} (\lambda_{-})^{1}_{( \beta_1} (\lambda_{-})^{2}_{\beta_2} (\lambda_{-})^{3}_{\beta_4 )}$. However $\epsilon^{\alpha_1 \alpha_2 \alpha_4} \epsilon_{\alpha_1 \alpha_2 \beta_3} \epsilon_{\alpha_4 \alpha_5 \beta_6} = 2 \epsilon_{\beta_3 \alpha_5 \beta_6}$ which is obviously antisymmetric in $\beta_3$ and $\beta_6$ and so would vanish when contacted with the symmetrized Wilson line indices.

Therefore, for the minimal choice we have $n$ in the range $0 \le n \le k/3 - 1$, giving fugacity $\prod_{n = 0}^{k/3 - 1} q^{2(n + 1)} = q^{\sum_{n = 0}^{k/3 - 1} 2(n + 1)} = q^{k(k+3)/9}$. As for the $SU(2)_{-2}$ case, for the non-minimal case we get cancellations from operators where we replace a derivative with a gaugino, and so there are no higher order contributions.

%proof
We now alternatively give an analytic proof directly from the half-index expression which can also be used to discuss Wilson lines in other representations. 
In this case we are interested in matrix integrals of the form
\begin{align}
I_{\alpha \beta} & = (q)_{\infty}^2 \oint \frac{ds_1}{2 \pi i s_1} \frac{ds_2}{2 \pi i s_2} \left( \prod_{1 \le i \le j \le 3} (s_i^{\pm} s_j^{\mp}; q)_{\infty} \right) s_1^{\alpha} s_2^{\beta}, 
\end{align}
where $\alpha, \beta \in \Zb$.

We now use the Jacobi triple product identity three times to get
\begin{align}
I_{\alpha \beta} = & \frac{1}{(q)_{\infty}} \sum_{l, m, n \in \Zb} (-1)^{l+m+n} \oint \frac{ds_1}{2 \pi i s_1} \frac{ds_2}{2 \pi i s_2} q^{\frac{1}{2} \left( l(l+1) + m(m+1) + n(n+1) \right)} \times
\nonumber \\
& \times s_1^{l + 2m + n + \alpha} s_2^{-l + m + 2n + \beta} \left( 1 - s_1 s_2^{-1} - s_1 s_2^2 + s_1^3 + s_1^3 s_2^3 - s_1^4 s_2^2 \right) \; .
\end{align}
Now let $\gamma = (\alpha + \beta)/3$ and define
\begin{align}
c_{\alpha \gamma} = & \frac{1}{(q)_{\infty}} \sum_{l, m, n \in \Zb} (-1)^{l+m+n} \oint \frac{ds_1}{2 \pi i s_1} \frac{ds_2}{2 \pi i s_2} q^{\frac{1}{2} \left( l(l+1) + m(m+1) + n(n+1) \right)} \times
\nonumber \\
& \times s_1^{l + 2m + n + \alpha} s_2^{-l + m + 2n + \beta} \; .
\end{align}
We see that the integrals pick up the contribution only for $l = -m + \gamma - \alpha$ and $n = -m -\gamma$, and so for integer $\gamma$ we have
\begin{align}
c_{\alpha \gamma} = & \frac{1}{(q)_{\infty}} (-1)^{\alpha} q^{\frac{1}{2}(-\alpha + \alpha^2 -2\alpha\gamma +2\gamma^2)} \sum_{m \in \Zb} (-1)^m q^{\frac{1}{2}(3m^2 - m -2m\alpha)}
\end{align}
while $c_{\alpha \gamma} = 0$ if $\gamma$ is not an integer.

We now evaluate this final sum using Euler's pentagonal number theorem
\begin{align}
\sum_{m \in \Zb} (-1)^m q^{\frac{1}{2}(3m^2 - m)} = (q)_{\infty} \; .
\end{align}
By shifting $m \rightarrow m - \alpha/3$, $m \rightarrow m - (\alpha - 1)/3$ or $m \rightarrow m - (\alpha + 1)/3$ for $\alpha \equiv 0, 1, -1 \mod 3$ respectively, the first two cases reduce the sum to Euler's pentagonal number theorem while for $\alpha \equiv -1 \mod 3$ the result is zero since
\begin{align}
    \sum_{m \in \Zb} (-1)^m q^{\frac{3}{2}m(m - 1)} = 0 \; .
\end{align}
due to the antisymmetry of the sum under $m \rightarrow 1-m$. The final result is
\begin{align}
c_{\alpha \gamma} = & \left \{ \begin{array}{rcl} (-1)^{\alpha/3} q^{-\frac{1}{6}\alpha(\alpha-1)} & , & \alpha = 0 \mod 3 \\ (-1)^{(\alpha-1)/3} q^{-\frac{1}{6}\alpha(\alpha-1)} & , & \alpha = 1 \mod 3 \\ 0 & , & \alpha = -1 \mod 3 \end{array} \right. \; .
\end{align}
We can then easily calculate
\begin{align}
\label{genSU3_k3}
I_{\alpha \beta} = & \left \{ \begin{array}{rcl} q^{\frac{1}{3}\alpha(\alpha - 1) + \gamma(\gamma - \alpha)} \left( 1 + q^{\frac{2}{3}\alpha - \gamma} + q^{-\frac{1}{3}\alpha + \gamma} + q^{\alpha - \gamma} + q^{\gamma} + q^{\frac{2}{3}\alpha} \right) & , & \alpha = 0 \mod 3 \\ - q^{\frac{1}{3}\alpha(\alpha - 1) + \gamma(\gamma - \alpha)} \left( 1 + q^{\alpha - \gamma} + q^{\gamma} \right) & , & \alpha = 1 \mod 3 \\ - q^{\frac{1}{3}\alpha(\alpha + 1) + \gamma(\gamma - \alpha - 1)} \left( 1 + q^{-\alpha + 2 \gamma} + q^{\gamma} \right) & , & \alpha = -1 \mod 3 \end{array} \right.
\end{align}
where we recall that this applies for $\gamma = (\alpha + \beta)/3 \in \Zb$, otherwise $I_{\alpha \beta} = 0$.

It is now straightforward to calculate the charge-$n$ Wilson line
\begin{align}
\langle W_n \rangle & = \frac{(q)_{\infty}^2}{3!} \oint \frac{ds_1}{2 \pi i s_1} \frac{ds_2}{2 \pi i s_2} \left( \prod_{1 \le i \le j \le 3} (s_i^{\pm} s_j^{\mp}; q)_{\infty} \right) 
  \left( s_1^n + s_2^n + s_1^{-n} s_2^{-n} \right)
  \nonumber \\
& = q^{\frac{n(n-3)}{9}} \left( 1 + q^{\frac{n}{3}} + q^{\frac{2n}{3}} \right)
\end{align}
for $n \equiv 0 \mod 3$ and the integral is zero otherwise.

These results can be used to derive the expression \eqref{su3_m3_sym} for rank-$k$ symmetric representation Wilson lines and indeed other representations although the calculations are in general lengthy so we do not present any details here.

%%%%%%%%%%%%%%%%%%%%%%%%%%%%%%%%%%
\subsubsection{Grand canonical one-point function}
\label{sec_1pt_su3G}
%%%%%%%%%%%%%%%%%%%%%%%%%%%%%%%%%%
We can consider the case where $n = 3k$ and again calculate the grand canonical ensemble by summing over the charge $n$, or equivalently over $k$
\begin{align}
    \sum_{k \in \Zb} \langle W_{3k} \rangle^{SU(3)_{-3}} \Lambda^{k}
     = & \Lambda^{1/2} \sum_{m = 0}^{2} q^{m/2 - 1/4} \vartheta_1(\lambda + m\tau + \frac{1}{2}; 2\tau) \; .
\end{align}

%%%%%%%%%%%%%%%%%%%%%%%%%%%%%%%%%%
\subsubsection{Two-point function}
\label{sec_2pt_su3sym}
%%%%%%%%%%%%%%%%%%%%%%%%%%%%%%%%%%
The two-point function of a pair of the Wilson lines transforming in the rank-$k$ symmetric representation is given by
\begin{align}
\label{su3_2pt_largesym}
\langle W_{(3l-m)} W_{\overline{(3l-m)}}\rangle^{SU(3)_{-3}}
&=\sum_{n=0}^{l+\delta_{m,0}}q^{3n^2+2n}-\sum_{n=1}^{l} q^{3n^2-2n}
\end{align}
where $l=1,2,\cdots$; $m=0,1,2$. 

In the large representation limit $k\rightarrow \infty$, 
the two-point function becomes an infinite series 
\begin{align}
\langle W_{(\infty)} W_{\overline{(\infty)}}\rangle^{SU(3)_{-3}}
&=\sum_{n=0}^{\infty}q^{3n^2+2n}-\sum_{n=1}^{\infty} q^{3n^2-2n}. 
\end{align}

%%%%%%%%%%%%%%%%%%%%%%%%%%%%%%%%%%
\subsection{$k=-7/2$}
\label{sec_su3_km7/2}
%%%%%%%%%%%%%%%%%%%%%%%%%%%%%%%%%%
Taking $N_f=1$, $N_a=0$ and $M=1$ with identification of the two global $U(1)$ fugacities, or equivalently taking a single fundamental chiral with Neumann boundary conditions, the $SU(3)$ Chern-Simons theory has level $k=-7/2$. 
The half-index is given by 
\begin{align}
\mathbb{II}_{\mathcal{N}}^{SU(3)_{-7/2}}(x;q)
&=\frac{(q;q)_{\infty}^2}{3!}\oint 
\frac{ds_1}{2\pi is_1}
\frac{ds_2}{2\pi is_2}
\prod_{1\le i< j\le3}
(s_i^{\pm}s_j^{\mp};q)_{\infty}
\prod_{i=1}^3 
(q^{\frac12}s_i x;q)_{\infty}. 
\end{align}
It has an expansion 
\begin{align}
\mathbb{II}_{\mathcal{N}}^{SU(3)_{-7/2}}
&=1-x^3q^{3/2}-x^3q^{5/2}-x^3q^{7/2}-x^3q^{9/2}-x^3q^{11/2}+x^6q^6-x^3q^{13/2}
\nonumber\\
&+x^6q^{7}-x^3q^{15/2}+2x^6q^8-x^3q^{17/2}+2x^6q^9-x^3q^{19/2}+\cdots
\end{align}
We find that it precisely agrees with
\begin{align}
\sum_{n=0}^{\infty}
\frac{(-1)^n q^{\frac{3n^2}{2}} x^{3n}}
{(1-q)(1-q^2) \cdots (1-q^n)}. 
\end{align}
These results can be understood in terms of the fundamental representation fermion in the Neumann chiral multiplet and its derivates which must be combined in the antisymmetric rank-$3$ representation to form a gauge singlet. Furthermore, no more than three such fermions can appear with the same number of derivatives otherwise the product will vanish due to antisymmetry.

%%%%%%%%%%%%%%%%%%%%%%%%%%%%%%%%%%
\subsubsection{One-point function}
\label{sec_1pt_su3km7/2}
%%%%%%%%%%%%%%%%%%%%%%%%%%%%%%%%%%
The one-point function of the Wilson line in the fundamental representation is evaluated as
\begin{align}
\langle W_1\rangle^{SU(3)_{-7/2}}(x;q)
&=x^2q-x^5q^{9/2}-x^5q^{11/2}-x^5q^{13/2}-x^5q^{15/2}-x^5q^{17/2}-x^5q^{19/2}
\nonumber\\
&-x^5q^{21/2}+x^8q^{11}-x^5q^{23/2}+x^8q^{12}-x^5q^{25/2}+2x^8q^{13}+\cdots
\end{align}
We find that it is given by
\begin{align}
\sum_{n=0}^{\infty}\frac{(-1)^n q^{\frac{3n^2}{2}+2n+1}x^{3n+2}}
{(1-q)(1-q^2)\cdots (1-q^n)}. 
\end{align}

%q-difference
The half-index and the one-point function satisfy the $q$-difference equation
\begin{align}
\label{qdiff_su3k72}
\mathbb{II}_{\mathcal{N}}^{SU(3)_{-7/2}}(q^{2/3}x;q)
&=q^{-1}x^{-2}\langle W_{1}\rangle^{SU(3)_{-7/2}}(x;q). 
\end{align}
This can be interpreted in terms of the duality with the dual of the Wilson line being a vortex line, but note that as for the similar $SU(2)_{-5/2}$ theory the $q$-difference equation involves a fractional power of $q$, in particular $q^{(N-1)/N}$ for $SU(N)_{-N-1}$.

%%%%%%%%%%%%%%%%%%%%%%%%%%%%%%%%%%
\subsection{$k=-4$}
\label{sec_su3_km4}
%%%%%%%%%%%%%%%%%%%%%%%%%%%%%%%%%%
When $N_f+N_a=2$ and $M=0$ or $N_f+N_a=0$ and $M=1$, 
we have the 3d $\mathcal{N}=2$ $SU(3)$ Chern-Simons theory with level $k=-4$. 
For $N_f=N_a=0$ and $M=1$, the half-index is given by the matrix integral
\begin{align}
\label{h_su3km4}
\mathbb{II}_{\mathcal{N}}^{SU(3)_{-4}}(x;q)
&=\frac{(q;q)_{\infty}^2}{3!}\oint 
\frac{ds_1}{2\pi is_1}
\frac{ds_2}{2\pi is_2}
\prod_{1\le i< j\le3}
(s_i^{\pm}s_j^{\mp};q)_{\infty}
\nonumber\\
&\times 
(q^{\frac12}s_1^{\pm} x^{\pm};q)_{\infty}
(q^{\frac12}s_2^{\pm} x^{\pm};q)_{\infty}
(q^{\frac12}s_1^{\mp}s_2^{\mp} x^{\pm};q)_{\infty}. 
\end{align}
It can be expanded as
\begin{align}
&1+q-(x^3+x^{-3})q^{3/2}+2q^2-(x^3+x^{-3})q^{5/2}+3q^3-2(x^3+x^{-3})q^{7/2}
\nonumber\\
&+5q^4-3(x^3+x^{-3})q^{9/2}+7q^5-5(x^3+x^{-3})q^{11/2}+(11+x^6+x^{-6})q^6+\cdots. 
\end{align}
This can be understood in terms of the gauge invariants which can formed from the fundamental fermion $\gamma$ and the antifundamental fermion $\bar{\gamma}$ in the 2d Fermi multiplet, and their derivatives. Note that for gauge group $SU(3)$ we can form a singlet from three fundamentals or three antifundamentals, while a fundamental (antifundamental) can be formed from two antifundamentals (fundamentals).

We find that the half-index is given by 
\begin{align}
\label{h_wzwU(1)_3}
\mathbb{II}_{\mathcal{N}}^{SU(3)_{-4}}(x;q)
&=\frac{1}{(q;q)_{\infty}}
\sum_{n\in \mathbb{Z}}(-1)^n q^{\frac{3n^2}{2}}x^{3n}. 
\end{align}
The expression (\ref{h_wzwU(1)_3}) agrees with 
the vacuum character of the $U(1)_{3}$ WZW model
as expected from the duality of boundary conditions proposed in \cite{Dimofte:2017tpi}
\begin{align}
&\textrm{$SU(3)_{-4}$ pure CS with Neumann b.c. $+$ fund. Fermi}
\nonumber\\
&\leftrightarrow 
\textrm{$U(1)_{3}$ pure CS with Dirichlet b.c.}
\end{align}

In the unflavored limit $x\rightarrow 1$, it agrees with
\begin{align}
\frac{\varphi(-q^{3/2})}{f(-q)}. 
\end{align}

%%%%%%%%%%%%%%%%%%%%%%%%%%%%%%%%%%
\subsubsection{One-point function}
\label{sec_1pt_su3km4}
%%%%%%%%%%%%%%%%%%%%%%%%%%%%%%%%%%
The one-point function of the Wilson line in the fundamental representation is expanded as
\begin{align}
\langle W_1\rangle^{SU(3)_{-4}}(x;q)
&=-x^{-1}q^{1/2}+x^2q-x^{-1}q^{3/2}+x^2q^2-2x^{-1}q^{5/2}+(2x^2+x^{-4})q^{3}
\nonumber\\
&-3x^{-1}q^{7/2}+(3x^2+x^{-4})q^4-(x^5+5x^{-1})q^{9/2}
+(5x^2+2x^{-4})q^5+\cdots. 
\end{align}
We find that it is given by
\begin{align}
\label{w1_su3km4}
\langle W_1\rangle^{SU(3)_{-4}}(x;q)
&=\frac{qx^2}{(q;q)_{\infty}}
\sum_{n\in \mathbb{Z}}
(-1)^n q^{\frac{3n^2}{2}+2n}x^{3n}
 = qx^2 \mathbb{II}_{\mathcal{N}}^{SU(3)_{-4}}(q^{2/3}x;q) \; . 
\end{align}
This can be interpreted in terms of the duality with the dual of the Wilson line being a vortex line, but note that as for the similar $SU(2)_{-3}$ theory the $q$-difference equation involves a fractional power of $q$, in particular $q^{(N-1)/N}$ for $SU(N)_{-N-1}$.

When we turn off the flavored fugacity $x\rightarrow 1$, it agrees with
\begin{align}
-q^{1/2}\frac{\psi(q^{3/2})}{\psi(q^{1/2})},
\end{align}
where $\psi(q)$ is Ramanujan's theta function (\ref{r_theta_psi}). 

The one-point function of the Wilson line of charge $2$ is obtained from 
the one-point function (\ref{w1_su3km4}) of the fundamental Wilson line upon replacing $x\rightarrow x^{-1}$ and changing the overall sign
\begin{align}
\label{w2_su3km4}
\langle W_2\rangle^{SU(3)_{-4}}(x;q)
&= - \frac{qx^{-2}}{(q;q)_{\infty}}
\sum_{n\in \mathbb{Z}}
(-1)^n q^{\frac{3n^2}{2}+2n}x^{-3n}. 
\end{align}

The half-index (\ref{h_su3km4}) and the one-point function (\ref{w2_su3km4}) obey the $q$-difference equation, following from (\ref{w1_su3km4})
\begin{align}
\label{qdiff_su3k4_w1}
\mathbb{II}_{\mathcal{N}}^{SU(3)_{-4}}(q^{2/3}x;q)
&= - q^{-1}x^{-2}\langle W_{2}\rangle^{SU(3)_{-4}}(x^{-1};q). 
\end{align}

By relabelling $n \to -n$ in (\ref{w2_su3km4}) we have
\begin{align}
\label{w2_su3km4_qdiff}
\langle W_2\rangle^{SU(3)_{-4}}(x;q)
&= - \frac{qx^{-2}}{(q;q)_{\infty}}
\sum_{n\in \mathbb{Z}}
(-1)^n q^{\frac{3n^2}{2}-2n}x^{3n} 
\end{align}
so we also have the $q$-difference equation
\begin{align}
\label{qdiff_su3k4_w2}
\mathbb{II}_{\mathcal{N}}^{SU(3)_{-4}}(q^{-2/3}x;q)
&= - q^{-1}x^{2}\langle W_{2}\rangle^{SU(3)_{-4}}(x;q). 
\end{align}

Furthermore, we find that the one-point functions of the Wilson line of larger charges are given by 
the half-index (\ref{h_su3km4}) and the one-point functions (\ref{w1_su3km4}) and (\ref{w2_su3km4}). 
For example, 
\begin{align}
\langle W_3\rangle^{SU(3)_{-4}}&=\mathbb{II}_{\mathcal{N}}^{SU(3)_{-4}},\\
\langle W_4\rangle^{SU(3)_{-4}}&=(1+q+q^2)\langle W_1\rangle^{SU(3)_{-4}},\\
\langle W_5\rangle^{SU(3)_{-4}}&=-q^3\langle W_2\rangle^{SU(3)_{-4}},\\
\langle W_6\rangle^{SU(3)_{-4}}&=-q^3\mathbb{II}_{\mathcal{N}}^{SU(3)_{-4}},\\
\langle W_7\rangle^{SU(3)_{-4}}&=-q^2\langle W_1\rangle^{SU(3)_{-4}},\\
\langle W_8\rangle^{SU(3)_{-4}}&=-q^3(1+q^2+q^4)\langle W_2\rangle^{SU(3)_{-4}},\\
\langle W_9\rangle^{SU(3)_{-4}}&=q^9 \mathbb{II}_{\mathcal{N}}^{SU(3)_{-4}}, \\
\langle W_{10}\rangle^{SU(3)_{-4}}&=-q^8\langle W_1\rangle^{SU(3)_{-4}},\\
\langle W_{11}\rangle^{SU(3)_{-4}}&=-q^7\langle W_2\rangle^{SU(3)_{-4}},\\
\langle W_{12}\rangle^{SU(3)_{-4}}&=q^9(1+q^3+q^6) \mathbb{II}_{\mathcal{N}}^{SU(3)_{-4}}. 
\end{align}

Note that Wilson lines of negative charge are related to those of positive charge since if we replace the integration variables (gauge fugacities) $s_i$ with $s_i^{-1}$ in the expression for $\langle W_{n} \rangle (x; q)$ we get the expression for $\langle W_{-n} \rangle (x^{-1}; q)$. For example we see that $\langle W_{-1} \rangle (x; q) = \langle W_{1} \rangle (x^{-1}; q) = - \langle W_{2} \rangle (x; q)$.

From this we can conjecture the general case of $\langle W_k \rangle^{SU(3)_{-4}}$ is given by the following:
\begin{itemize}
    \item A factor $\left\{ \begin{array}{rl} \mathbb{II}_{\mathcal{N}}^{SU(3)_{-4}}, & k \equiv 0 \mod 3 \\ \langle W_1\rangle^{SU(3)_{-4}}, & k \equiv 1 \mod 3 \\ \langle W_2\rangle^{SU(3)_{-4}}, & k \equiv 2 \mod 3 \end{array} \right.$
    \item A factor $(1 + q^{k/4} + q^{k/2})$ for $k \equiv 0 \mod 4$ for $k \ne 0$
    %\item A factor $-1$ for $k \equiv 5, 6, 7, 8, 10 , 11 \mod 12$
\end{itemize}
and a factor of a single power of $q$ and a sign $\pm1$ which we haven't determined but can be deduced from (\ref{genSU3_k3}).

%%%%%%%%%%%%%%%%%%%%%%%%%%%%%%%%%%
\subsection{$k = - 5/2 - M$}
\label{sec_su3_km5/2M}
%%%%%%%%%%%%%%%%%%%%%%%%%%%%%%%%%%
We expect the Neumann half-indices are equal to half-indices of dual $U(M)_{5/2 + M, 5/2}$ theories with Dirichlet boundary conditions for the gauge multiplet and a 3d fundamental chiral. We have already described the case of $M = 1$ where we identified the Fermi and 3d chiral $U(1)$ fugacities, equivalent to considering instead the case of a singe 3d fundamental chiral with Dirichlet boundary conditions. Here we list some results for $M = 2$ and $M = 3$ but the one-point functions of Wilson lines operators are more complicated than in the case of integer Chern-Simons level and we don't have any general conjectures or $q$-difference equations relating these one-point functions to the half-index in the absence of Wilson lines.

%%%%%%%%%%%%%%%%%%%%%%%%%%%%%%%%%%
\subsubsection{$k=-9/2$}
\label{sec_su3_km9/2}
%%%%%%%%%%%%%%%%%%%%%%%%%%%%%%%%%%
For the Chern-Simons level $k=-9/2$, 
we take two fundamental Fermi multiplets and a single fundamental chiral multiplet with the Neumann boundary condition. 
The half-index is given by 
\begin{align}
\mathbb{II}_{\mathcal{N}}^{SU(3)_{-9/2}}(x_{\alpha};q)
&=\frac{(q;q)_{\infty}^2}{3!}\oint 
\frac{ds_1}{2\pi is_1}
\frac{ds_2}{2\pi is_2}
\prod_{1\le i< j\le3}
(s_i^{\pm}s_j^{\mp};q)_{\infty}
\prod_{i=1}^{3}
\prod_{\alpha=1}^{2}
\frac{
(q^{\frac12}s_i^{\pm}x_{\alpha}^{\pm};q)_{\infty}
}{(q^{\frac12}s_i x_3;q)_{\infty}}. 
\end{align}
It has an expansion 
\begin{align}
\mathbb{II}_{\mathcal{N}}^{SU(3)_{-9/2}}
&=1+\left(
2+\frac{x_1}{x_2}+\frac{x_2}{x_1}-\frac{x_3}{x_1}-\frac{x_3}{x_2}
\right)q+
\Bigl(
x_1^2x_3+x_2^2x_3+x_1x_2x_3
\nonumber\\
&-x_1^3-x_1^{-3}-x_2^3-x_2^{-3}-x_1^{-1}x_2^{-2}-x_1^{-2}x_2^{-1}-x_1x_2^2-x_1^2x_2
\Bigr)q^{3/2}+\cdots. 
\end{align}
This result is easily interpreted in terms of the two fundamental fermions $\gamma^i$ and antifundamental fermions $\bar{\gamma}_i$from the two Fermi multiplets, along with the fundamental scalar $\phi$ from the 3d chiral with Neumann boundary conditions.

We find that it precisely coincides with
\begin{align}
\label{h_dualsu3km92}
\frac{1}{(q;q)_{\infty}^2}
\sum_{m_1,m_2\in \mathbb{Z}}
\frac{(-1)^{m_1+m_2} q^{\frac32 (m_1^2+m_2^2)}x_1^{3m_1} x_2^{3m_2}
(q^{1-m_{1}}x_1^{-1}x_3;q)_{\infty}
(q^{1-m_{2}}x_2^{-1}x_3;q)_{\infty}
}
{(q^{1+m_1-m_2}\frac{x_1}{x_2};q)_{\infty} (q^{1+m_2-m_1}\frac{x_2}{x_1};q)_{\infty}}. 
\end{align}
The series expression (\ref{h_dualsu3km92}) demonstrates that the dual description is indeed given by 
the $U(2)_{9/2, 5/2}$ Chern-Simons theory with a fundamental chiral with Dirichlet boundary conditions.

%%%%%%%%%%%%%%%%%%%%%%%%%%%%%%%%%%
\subsubsection{$k=-11/2$}
\label{sec_su3_km11/2}
%%%%%%%%%%%%%%%%%%%%%%%%%%%%%%%%%%
Taking two fundamental Fermi multiplets and a single fundamental chiral multiplet with the Neumann boundary condition, 
the Chern-Simons level $k=-11/2$ is allowed. 
The half-index reads
\begin{align}
\label{h_su3km112}
\mathbb{II}_{\mathcal{N}}^{SU(3)_{-11/2}}(x_{\alpha};q)
&=\frac{(q;q)_{\infty}^2}{3!}\oint 
\frac{ds_1}{2\pi is_1}
\frac{ds_2}{2\pi is_2}
\prod_{1\le i< j\le3}
(s_i^{\pm}s_j^{\mp};q)_{\infty}
\prod_{i=1}^{3}
\prod_{\alpha=1}^{3}
\frac{
(q^{\frac12}s_i^{\pm}x_{\alpha}^{\pm};q)_{\infty}
}{(q^{\frac12}s_i x_4;q)_{\infty}}. 
\end{align}
We find that it precisely coincides with
\begin{align}
\label{h_dualsu3km112}
\frac{1}{(q;q)_{\infty}^3}
\sum_{m_1,m_2,m_3\in \mathbb{Z}}
\frac{
\prod_{\alpha=1}^3
(-1)^{m_{\alpha}} q^{\frac32m_{\alpha}^2}
x_{\alpha}^{3m_{\alpha}}
(q^{1-m_{\alpha}}x_{\alpha}^{-1}x_4;q)_{\infty}
}
{
\prod_{\alpha<\beta}^{3}
(q^{1+m_{\alpha}-m_{\beta}} \frac{x_{\alpha}}{x_{\beta}};q)_{\infty} (q^{1+m_{\beta}-m_{\alpha}}\frac{x_{\beta}}{x_{\alpha}};q)_{\infty}}. 
\end{align}
The series expression (\ref{h_dualsu3km112}) demonstrates that the dual description is given by 
the $U(3)_{11/2, 5/2}$ Chern-Simons theory with a fundamental 3d chiral having Dirchlet boundary conditions.

%%%%%%%%%%%%%%%%%%%%%%%%%%%%%%%%%%
\subsection{$k = -3 - M$}
\label{sec_su3_km3M}
%%%%%%%%%%%%%%%%%%%%%%%%%%%%%%%%%%
We expect the Neumann half-indices are equal to half-indices of dual $U(M)_{3 + M, 3}$ theories with Dirichlet boundary conditions for the gauge multiplet and a 3d fundamental chiral. We have already described the case of $M = 1$ and here we list some further examples
noting consistency with the duality of boundary conditions proposed in \cite{Dimofte:2017tpi}
\begin{align}
&\textrm{$SU(N)_{-3 - M}$ pure CS with Neumann b.c. $+$ $M$ fund. Fermis}
\nonumber\\
&\leftrightarrow 
\textrm{$U(M)_{3 + M,3}$ pure CS with Dirichlet b.c.}
\end{align}

%%%%%%%%%%%%%%%%%%%%%%%%%%%%%%%%%%
\subsubsection{$k=-5$}
\label{sec_su3_km5}
%%%%%%%%%%%%%%%%%%%%%%%%%%%%%%%%%%
When $M=2$, we have the pure Chern-Simons theory with level $k=-5$. 
The half-index is 
\begin{align}
\label{h_su3km5}
\mathbb{II}_{\mathcal{N}}^{SU(3)_{-5}}(x_{\alpha};q)
&=\frac{(q;q)_{\infty}^2}{3!}\oint 
\frac{ds_1}{2\pi is_1}
\frac{ds_2}{2\pi is_2}
\prod_{1\le i< j\le3}
(s_i^{\pm}s_j^{\mp};q)_{\infty}
\nonumber\\
&\times 
\prod_{\alpha=1}^{2}
(q^{\frac12}s_1^{\pm} x_{\alpha}^{\pm};q)_{\infty}
(q^{\frac12}s_2^{\pm} x_{\alpha}^{\pm};q)_{\infty}
(q^{\frac12}s_1^{\mp}s_2^{\mp} x_{\alpha}^{\pm};q)_{\infty}. 
\end{align}
It agrees with 
the vacuum character of the $U(2)_{3}$ WZW model
\begin{align}
\label{h_wzwU(2)_3}
\frac{1}{(q;q)_{\infty}^2}
\sum_{m_1,m_2\in \mathbb{Z}}
\frac{(-1)^{m_1+m_2} q^{\frac32 (m_1^2+m_2^2)}x_1^{3m_1} x_2^{3m_2}}
{(q^{1+m_1-m_2}\frac{x_1}{x_2};q)_{\infty} (q^{1+m_2-m_1}\frac{x_2}{x_1};q)_{\infty}}. 
\end{align}

%%%%%%%%%%%%%%%%%%%%%%%%%%%%%%%%%%
\subsubsection{$k=-5$ One-point function}
\label{sec_1pt_su3km5}
%%%%%%%%%%%%%%%%%%%%%%%%%%%%%%%%%%
Similarly, we find that the one-point function of the fundamental Wilson line is given by
\begin{align}
\label{w1_su3km5}
&
\langle W_1\rangle^{SU(3)_{-5}}(x_{\alpha};q)
\nonumber\\
&=\frac{q}{(q;q)_{\infty}^2}
\sum_{m_1,m_2\in \mathbb{Z}}
\frac{(-1)^{m_1+m_2} q^{\frac32 (m_1^2+m_2^2)} x_1^{3m_1} x_2^{3m_2} 
(q^{2m_1}x_1^2+q^{2m_2}x_2^2+q^{m_1+m_2}x_1x_2)}
{(q^{1+m_1-m_2}\frac{x_1}{x_2};q)_{\infty}(q^{1+m_2-m_1}\frac{x_2}{x_1};q)_{\infty}}. 
\end{align}
We have the $q$-difference equation 
\begin{align}
&
\langle W_1\rangle^{SU(3)_{-5}}(x_1,x_2;q)
\nonumber\\
&=qx_1^2 \mathbb{II}_{\mathcal{N}}^{SU(3)_{-5}}(q^{2/3}x_1,x_2;q)
+qx_2^2 \mathbb{II}_{\mathcal{N}}^{SU(3)_{-5}}(x_1,q^{2/3}x_2;q)
\nonumber\\
&+qx_1 x_2 \mathbb{II}_{\mathcal{N}}^{SU(3)_{-5}}(q^{1/3}x_1,q^{1/3}x_2;q). 
\end{align}

%%%%%%%%%%%%%%%%%%%%%%%%%%%%%%%%%%
\subsubsection{$k=-6$}
\label{sec_su3_km6}
%%%%%%%%%%%%%%%%%%%%%%%%%%%%%%%%%%
For $M=3$, we have the pure Chern-Simons theory with level $k=-6$. 
We have the half-index
\begin{align}
\label{h_su3km6}
\mathbb{II}_{\mathcal{N}}^{SU(3)_{-6}}(x_{\alpha};q)
&=\frac{(q;q)_{\infty}^2}{3!}\oint 
\frac{ds_1}{2\pi is_1}
\frac{ds_2}{2\pi is_2}
\prod_{1\le i< j\le3}
(s_i^{\pm}s_j^{\mp};q)_{\infty}
\nonumber\\
&\times 
\prod_{\alpha=1}^{3}
(q^{\frac12}s_1^{\pm} x_{\alpha}^{\pm};q)_{\infty}
(q^{\frac12}s_2^{\pm} x_{\alpha}^{\pm};q)_{\infty}
(q^{\frac12}s_1^{\mp}s_2^{\mp} x_{\alpha}^{\pm};q)_{\infty}. 
\end{align}
It matches with 
the vacuum character of the $U(3)_{3}$ WZW model 
\begin{align}
\label{h_wzwU(3)_3}
\frac{1}{(q;q)_{\infty}^3}
\sum_{m_1,m_2,m_3\in \mathbb{Z}}
\frac{(-1)^{m_1+m_2+m_3} q^{\frac32 (m_1^2+m_2^2+m_3^2)}x_1^{3m_1} x_2^{3m_2} x_3^{3m_3}}
{\prod_{\alpha<\beta}^{3} (q^{1+m_{\alpha}-m_{\beta}}\frac{x_{\alpha}}{x_{\beta}};q)_{\infty} 
(q^{1+m_{\beta}-m_{\alpha}}\frac{x_{\beta}}{x_{\alpha}};q)_{\infty}}. 
\end{align}

%%%%%%%%%%%%%%%%%%%%%%%%%%%%%%%%%%
\subsubsection{$k=-6$ One-point function}
\label{sec_1pt_su3km6}
%%%%%%%%%%%%%%%%%%%%%%%%%%%%%%%%%%

We find that 
the one-point function of the fundamental Wilson line is given by
\begin{align}
\label{w1_su3km6}
&
\langle W_1\rangle^{SU(3)_{-6}}(x_{\alpha};q)
\nonumber\\
&=\frac{q}{(q;q)_{\infty}^3}
\sum_{m_1,m_2,m_3\in \mathbb{Z}}
\frac{(-1)^{m_1+m_2+m_3} q^{\frac32 (m_1^2+m_2^2+m_3^2)}x_1^{3m_1} x_2^{3m_2} x_3^{3m_3}}
{\prod_{\alpha<\beta}^{3} (q^{1+m_{\alpha}-m_{\beta}}\frac{x_{\alpha}}{x_{\beta}};q)_{\infty} 
(q^{1+m_{\beta}-m_{\alpha}}\frac{x_{\beta}}{x_{\alpha}};q)_{\infty}}
\nonumber\\
&\times 
(q^{2m_1}x_1^2+q^{2m_2}x_2^2+q^{2m_3}x_3^2
+q^{m_1+m_2}x_1x_2
+q^{m_1+m_3}x_1x_3
+q^{m_2+m_3}x_2x_3
). 
\end{align}

The $q$-difference equation satisfied by 
the half-index (\ref{h_su3km6}) and the one-point function (\ref{w1_su3km6}) is
\begin{align}
&
\langle W_1\rangle^{SU(3)_{-6}}(x_1,x_2,x_3;q)
\nonumber\\
&=qx_1^2 \mathbb{II}_{\mathcal{N}}^{SU(3)_{-6}}(q^{2/3}x_1,x_2,x_3;q)
+qx_2^2 \mathbb{II}_{\mathcal{N}}^{SU(3)_{-6}}(x_1,q^{2/3}x_2,x_3;q)
\nonumber\\
&+qx_3^2 \mathbb{II}_{\mathcal{N}}^{SU(3)_{-6}}(x_1,x_2,q^{2/3}x_3;q)
+qx_1x_2 \mathbb{II}_{\mathcal{N}}^{SU(3)_{-6}}(q^{1/3}x_1,q^{1/3}x_2,x_3;q)
\nonumber\\
&+qx_1x_3 \mathbb{II}_{\mathcal{N}}^{SU(3)_{-6}}(q^{1/3}x_1,x_2,q^{1/3}x_3;q)
+qx_2x_3 \mathbb{II}_{\mathcal{N}}^{SU(3)_{-6}}(x_1,q^{1/3}x_2,q^{1/3}x_3;q). 
\end{align}

%%%%%%%%%%%%%%%%%%%%%%%%%%%%%%%%%%
%%%%%%%%%%%%%%%%%%%%%%%%%%%%%%%%%%
\section{$SU(N)$}
\label{sec_suN}
%%%%%%%%%%%%%%%%%%%%%%%%%%%%%%%%%%
%%%%%%%%%%%%%%%%%%%%%%%%%%%%%%%%%%
In agreement with the analytic results for $SU(2)$ and $SU(3)$,
we propose conjectures for general $SU(N)$ gauge group. 
We have numerically confirmed them for higher $N$. 

%%%%%%%%%%%%%%%%%%%%%%%%%%%%%%%%%%
\subsection{$k=-N$}
\label{sec_suN_kmN}
%%%%%%%%%%%%%%%%%%%%%%%%%%%%%%%%%%
When $N_f=0$ and $M=0$, we have the 3d $\mathcal{N}=2$ $SU(N)$ pure Chern-Simons theory with level $k=-N$. 
When the vector multiplet obeys the Neumann boundary condition in the absence of the Wilson line, 
there is no gauge invariant BPS local operator which contributes to the half-index. 

%%%%%%%%%%%%%%%%%%%%%%%%%%%%%%%%%%
\subsubsection{One-point function}
\label{sec_1pt_suN_kmN}
%%%%%%%%%%%%%%%%%%%%%%%%%%%%%%%%%%
We conjecture that the one point function for a charge $n$ Wilson line in the 3d $\mathcal{N} = 2$ $SU(N)$ pure Chern-Simons theory is given by
\begin{align}
    &\langle W_n \rangle^{SU(N)_{-N}} (q)
    \nonumber \\
    &=\frac{(q;q)_{\infty}^{N-1}}{N!} \oint \left( \prod_{i = 1}^{N-1} \frac{ds_i}{2 \pi i s_i} \right) \left( \prod_{i \ne j}^N (s_i s_j^{-1}; q)_{\infty} \right) \left( s_1^n + s_2^n + \cdots + s_N^n \right)
    \nonumber \\
    & = \left \{
    \begin{array}{lcl}
        (-1)^{k(N-1)} \sum_{l = 0}^{N-1} q^{(N-1)k(k-1)/2 + lk} = \frac{(-1)^{k(N-1)} q^{(N-1)k(k-1)/2}(1 - q^{kN})}{1 - q^k} & , & n = kN
        \\
        0 & , & n \ne kN
    \end{array}
    \right .
\end{align}
where $k, N \in \Zb$ noting that $s_N = (s_1 s_2 \cdots s_{N-1})^{-1}$. It is straightforward to see the zero result in the case that $n$ is not a multiple of $N$.

Also we conjecture that 
the one-point function of the Wilson line $W_{(kN)}$ transforming in the rank-$kN$ symmetric representation is given by
\begin{align}
W_{(kN)}^{SU(N)_{-N}}&=(-1)^{(N-1)k} q^{\frac{N-1}{2}k (k+1)}. 
\end{align}

This result follows from the same argument used for $SU(2)_{-2}$ and in particular $SU(3)_{-3}$. Here we group $N-1$ operators $D_z^n \lambda_{-}$ to create an operator with $N$ fundamental indices so we can contract with $kN$ synmmetrized indices from a Wilson line in the rank-$kN$ symmetric representation by taking $k$ such groups. The minimal (non-vanishing) case is when the groups have distinct $n$ in the range $0 \le n \le k-1$ giving fugacity $\prod_{n=0}^{k-1} (-1)^{N-1} q^{(N-1)(n+1)} = (-1)^{(N-1)k} q^{\frac{N-1}{2}k (k+1)}$.

%%%%%%%%%%%%%%%%%%%%%%%%%%%%%%%%%%
\subsubsection{Grand canonical one-point function}
\label{sec_1pt_suN_kmN_G}
%%%%%%%%%%%%%%%%%%%%%%%%%%%%%%%%%%
Now, consider the case where $n$ is a multiple of $N$, and calculate the grand canonical ensemble by summing over the charge $n$, or equivalently over $k$
\begin{align}
    \sum_{k \in \Zb} \langle W_{kN} \rangle^{SU(N)_{-N}} \Lambda^{k} = &
    (-1)^{N-1} \Lambda q^{-(N-1)/2} \mathcal{A}_{N-1}(\tau, \lambda; \tau)
     - \mathcal{A}_{N-1}(0, \lambda + \tau; \tau)
     \nonumber \\
     = & i^{N+1} \Lambda^{1/2} \sum_{m = 0}^{N-1} q^{m/2 - (N-1)/8} \vartheta_1(\lambda + m\tau + \frac{N-2}{2}; (N-1)\tau) \; ,
\end{align}
where for the first line we note that the first order, level $k$ Appell-Lerch sums can be defined as
\begin{align}
\label{app1a}
\mathcal{A}_{k}(u, v; \tau)&
=U^{k/2} \sum_{n \in \mathbb{Z}} \frac{(-1)^{kn} q^{kn(n+1)/2} V^n}{1 - Uq^n} \; 
\end{align}
where we have denoted $q = \exp(2\pi i \tau)$, $U = \exp(2\pi i u)$ and $V = \exp(2\pi i v)$. The Appell-Lerch sums are holomorphic but not modular. However, they have modular completions which are the weight $1$ Jacobi forms
\cite{ANDREWS199160, https://doi.org/10.1112/plms/pdu007}
\begin{align}
\label{app1b}
\hat{\mathcal{A}}_{k}(u, v; \tau)&=
\mathcal{A}_{k}(u, v; \tau) + \mathcal{R}_{k}(u, v; \tau)
\end{align}
where
\begin{align}
\label{app1c}
\mathcal{R}_{k}(u, v; \tau)&
=\frac{-i}{2k}U^{(k - 1)/2} \sum_{m = 0}^{k-1}
\vartheta_{1}\left( \frac{v+m}{k} + \frac{(k-1)\tau}{2k} ; \frac{\tau}{k} \right) \nonumber \\
 &\times R\left( u - \frac{v+m}{k} - \frac{(k-1)\tau}{2k} ;
 \frac{\tau}{k} \right) \\
\label{app1d}
R(w ; \tau)&=
\sum_{\nu \in \mathbb{Z} + 1/2}
\left( \mathrm{sgn}(\nu) - \mathrm{Erf}
\left( \sqrt{2\pi\tau_2}\left( \nu + \frac{\Im(w)}{\tau_2} \right) \right) \right) \nonumber \\
 &\times (-1)^{\nu-1/2}W^{-\nu}q^{-\nu^2/2}
\end{align}
and $\tau_2 = \Im(\tau)$.
These functions and higher order Appell-Lerch sums \cite{Zwegers:2008zna} have appeared in relation to black hole physics \cite{Dabholkar:2012nd, Ashok:2014nua} and supergroup WZW models arising as topologically tgwisted theories at the intersection of M2- and M5-branes \cite{Okazaki:2016pne, Okazaki:2015fiq}.

It is interesting to note that using the elliptic transformation properties of $\hat{\mathcal{A}}_{N-1}$ there is an identity
\begin{align}
    (-1)^{N-1} \Lambda q^{-(N-1)/2} \hat{\mathcal{A}}_{N-1}(\tau, \lambda; \tau) - \hat{\mathcal{A}}_{N-1}(0, \lambda + \tau; \tau) & = 0
\end{align}
so we can also write
\begin{align}
    \sum_{k \in \Zb} \langle W_{kN} \rangle^{SU(N)_{-N}} \Lambda^{k} = &
    (-1)^{N} \Lambda q^{-(N-1)/2} \mathcal{R}_{N-1}(\tau, \lambda; \tau)
     + \mathcal{R}_{N-1}(0, \lambda + \tau; \tau) \; .
\end{align}

%%%%%%%%%%%%%%%%%%%%%%%%%%%%%%%%%%
\subsubsection{Two-point function}
\label{sec_2pt_suN_kmN}
%%%%%%%%%%%%%%%%%%%%%%%%%%%%%%%%%%

We conjecture that 
the two-point function of a pair of Wilson line operators in the large symmetric representation is given by
\begin{align}
\label{largerep_suN_N2pt}
\langle W_{(\infty)} W_{\overline{(\infty)}}\rangle^{SU(N)_{-N}}
&=\Psi(q^{2N-1},q)
\end{align}
where $\Psi(a,b)$ is the false general Ramanujan's theta function (\ref{r_theta_Psi}). 
In addition to the result (\ref{su2_2pt_largesym}) for $SU(2)$ and (\ref{su3_2pt_largesym}) for $SU(3)$, 
we have checked that (\ref{largerep_suN_N2pt}) is correct for $SU(4)$ and $SU(5)$ by expanding the large symmetric Wilson line two-point functions up to order $q^{10}$. 
The R.H.S. of (\ref{largerep_suN_N2pt}) is called the false general Ramanujan's theta function of order $N$ \cite{rogers1917two}. 
The identities of false general Ramanujan's theta functions arise from the Bailey pairs \cite{MR25025}. 
For example, for $SU(2)$ we have \cite{MR49225,MR4336217}
\begin{align}
\langle W_{(\infty)} W_{\overline{(\infty)}}\rangle^{SU(2)_{-2}}
&=
\Psi(q^3,q)=
\sum_{n=0}^{\infty}
\frac{(-1)^{n^2}q^{n(n+1)}(q;q^2)_{n}}
{(-q;q^2)_{n+1}(-q^2;q^2)_{n}}. 
\end{align}
For $N=3$ we have \cite{MR25025}
\begin{align}
\langle W_{(\infty)} W_{\overline{(\infty)}}\rangle^{SU(3)_{-3}}
&=
\Psi(q^5,q)=
\sum_{n=0}^{\infty}
\frac{(-1)^n q^{3n(n+1/2}(q;q)_{3n+1}}{(q^3;q^3)_{2n+1}}. 
\end{align}
For $SU(4)$ it follows that \cite{MR2135178}
\begin{align}
\langle W_{(\infty)} W_{\overline{(\infty)}}\rangle^{SU(4)_{-4}}
&=
\Psi(q^7,q)=
\sum_{n=0}^{\infty}
\frac{(-1)^n q^{2n(n+1)} (q^4;q^4)_{n} (q;q^2)_{2n+1}}
{(q^4;q^4)_{2n+1}}. 
\end{align}
For $SU(5)$ one finds \cite{rogers1917two}
\begin{align}
\langle W_{(\infty)} W_{\overline{(\infty)}}\rangle^{SU(5)_{-5}}
&=\Psi(q^9,q)=\sum_{n=0}^{\infty}
\frac{(-1)^{n(n+1)/2} q^{n(n+3}/2}
{(-q;q^2)_{n+1}}
\end{align}

For general $N$ it can be also expressed as \cite{MR3485997}
\begin{align}
\langle W_{(\infty)} W_{\overline{(\infty)}}\rangle^{SU(N)_{-N}}
&=(q)_{\infty}
\sum_{l_1=0}^{\infty}\cdots \sum_{l_{N-1}=0}^{\infty}
\frac{q^{\sum_{j=1}^{N-1} i_{j} (i_j+1)}}
{(q)_{l_{N-1}}^{2} \prod_{j=1}^{N-2}(q)_{l_j}}
\end{align}
where $i_j=\sum_{s=j}^{N-1}l_s$. 
It would be intriguing to figure out the physical meaning of such series expressions of the large representation two-point functions. 

%%%%%%%%%%%%%%%%%%%%%%%%%%%%%%%%%%
\subsection{$k=-N-1/2$}
\label{sec_suN_kmN1/2}
%%%%%%%%%%%%%%%%%%%%%%%%%%%%%%%%%%
For $N_f=1$, $N_a=0$ and $M=1$, or with just a single 3d fundamental chiral with Dirichlet boundary conditions,
we have the $SU(N)$ Chern-Simons theory with level $k=-N-1/2$. 
For the latter theory, having a singe Dirichlet chiral, or for the former theory with a specialisation of the global $U(1)$ fugacities, for $N=4$, $5$ the half-indices are expanded as
\begin{align}
\mathbb{II}_{\mathcal{N}}^{SU(4)_{-9/2}}(x;q)
&=1+x^4q^2+x^4q^3+x^4q^4+x^4q^5+x^4q^7+(x^4+x^8)q^8
\nonumber\\
&+(x^4+x^8)q^9+(x^4+2x^8)q^{10}+\cdots.
, \\
\mathbb{II}_{\mathcal{N}}^{SU(5)_{-11/2}}(x;q)
&=1-x^5q^{5/2}-x^5q^{7/2}-x^5q^{9/2}
+\cdots
\end{align}

%\textcolor{red}{Discuss dual theory here and for one-point function below, noting that the expressions are simplest for a Dirichlet chiral rather than Neumann chiral plus Fermi.}

We conjecture that the half-index is given by
\begin{align}
\mathbb{II}_{\mathcal{N}}^{SU(N)_{-N-1/2}}(x;q)
&=\sum_{n=0}^{\infty}
\frac{(-1)^{Nn}q^{\frac{Nn^2}{2}}x^{Nn}}
{(1-q)(1-q^2)\cdots (1-q^n)}. 
\end{align}
which is consistent with a dual $U(1)_{N-1/2}$ theory with a single Dirichlet chiral after identification of the $U(1)$ global flavor symmetry with the $U(1)$ global symmetry arising fom the Dirichlet boundary condition for the vector multiplet.

%%%%%%%%%%%%%%%%%%%%%%%%%%%%%%%%%%
\subsubsection{One-point function}
\label{sec_1pt_suN_kmN1/2}
%%%%%%%%%%%%%%%%%%%%%%%%%%%%%%%%%%
For example, for $N=4$ we find that the one-point function has an expansion
\begin{align}
\langle W_{1}\rangle^{SU(4)_{-9/2}}(x;q)
&=-x^3q^{3/2}-x^7q^{13/2}-x^7q^{15/2}-x^7q^{17/2}-x^7q^{19/2}
+\cdots.
%\\
%\langle W_{1}\rangle^{SU(5)_{-11/2}}
%&=x^4q^2
%+\cdots. 
\end{align}

It precisely agrees with
\begin{align}
-\sum_{n=0}^{\infty}\frac{q^{2n^2+3n+\frac32} x^{4n+3}}
{(1-q)(1-q^2)\cdots (1-q^n)}.  
%\\
%\sum_{n=0}^{\infty}\frac{
%(-1)^n q^{\frac{5n^2}{2}+4n+2} x^{5n+4}}
%{(1-q)(1-q^2)\cdots (1-q^n)}. 
\end{align}

We are led to propose an expression of the one-point function for general $N$
\begin{align}
\langle W_{1}\rangle^{SU(N)_{-N-1/2}}(x;q)
&=\sum_{n=0}^{\infty}\frac{(-1)^{Nn+(N-1)} q^{\frac{Nn^2}{2}+(N-1)n+\frac{N-1}{2}} x^{Nn+(N-1)}}
{(1-q)(1-q^2)\cdots (1-q^n)}. 
\end{align}

Consequently, 
we find the $q$-difference equation
\begin{align}
\mathbb{II}_{\mathcal{N}}^{SU(N)_{-N-1/2}}(q^{(N-1)/N}x;q)
&=(-1)^{N-1}q^{-(N-1)/2}x^{-(N-1)}\langle W_{1}\rangle^{SU(N)_{-N-1/2}}(x;q). 
\end{align}

%%%%%%%%%%%%%%%%%%%%%%%%%%%%%%%%%%
\subsection{$k=-N-1$}
\label{sec_suN_kmN+1}
%%%%%%%%%%%%%%%%%%%%%%%%%%%%%%%%%%

For $N=4$, $5$ the half-indices are evaluated as
\begin{align}
\mathbb{II}_{\mathcal{N}}^{SU(4)_{-5}}(x;q)
&=1+q+(2+x^4+x^{-4})q^2+(3+x^4+x^{-4})q^3+(5+2x^4+2x^{-4})q^4
\nonumber\\
&+(7+3x^4+3x^{-4})q^5+\cdots, \\
\mathbb{II}_{\mathcal{N}}^{SU(5)_{-6}}(x;q)
&=1+q+2q^2-(x^5+x^{-5})q^{5/2}+3q^3-(x^5+x^{-5})q^{7/2}+5q^4
\nonumber\\
&-(2x^5+2x^{-5})q^{9/2}+7q^5+\cdots.
\end{align}

We conjecture that, consistent with a dual $U(1)_{N}$ theory with no matter content, the half-index is given by
\begin{align}
\label{h_suNkmN+1}
\mathbb{II}_{\mathcal{N}}^{SU(N)_{-N-1}}(x;q)
&=\frac{1}{(q;q)_{\infty}}
\sum_{n\in \mathbb{Z}}(-1)^{Nn}q^{\frac{Nn^2}{2}}x^{Nn}, 
\end{align}
which becomes 
\begin{align}
\frac{\varphi((-q^{1/2})^{N})}{f(-q)}
\end{align}
in the unflavored limit. 

%\textcolor{red}{[NOTE] Physical interpretation}

%\textcolor{red}{[NOTE] Dual Dirichlet? }

%%%%%%%%%%%%%%%%%%%%%%%%%%%%%%%%%%
\subsubsection{One-point function}
\label{sec_1pt_suN_kmN+1}
%%%%%%%%%%%%%%%%%%%%%%%%%%%%%%%%%%

For $N=4$, $5$ the one-point functions can be expanded as
\begin{align}
\langle W_{1}\rangle^{SU(4)_{-5}}
&=-x^{-1}q^{1/2}-(x^3+x^{-1})q^{3/2}-(x^3+2x^{-1})q^{5/2}-(2x^3+3x^{-1}+x^{-5})q^{7/2}
\nonumber\\
&-(3x^3+5x^{-1}+x^{-5})q^{9/2}-(5x^3+7x^{-1}+2x^{-5})q^{11/2}+\cdots,\\
\langle W_{1}\rangle^{SU(5)_{-6}}
&=-x^{-1}q^{1/2}-x^{-1}q^{3/2}+x^4q^2-2x^{-1}q^{5/2}+x^4q^3-3x^{-1}q^{7/2}
\nonumber\\
&+(2x^4+x^{-6})q^{4}-5x^{-1}q^{9/2}+(3x^4+x^{-6})q^5+\cdots. 
\end{align}
We find that they precisely coincide with 
\begin{align}
&-\frac{q^{\frac32}x^3}{(q;q)_{\infty}}
\sum_{n\in \mathbb{Z}}q^{2n^2+3n}x^{4n}, \\
&\frac{q^2 x^4}{(q;q)_{\infty}}
\sum_{n\in \mathbb{Z}}(-1)^n q^{\frac{5n^2}{2}+4n}x^{5n}, 
\end{align}
respectively. 
In particular, the unflavored one-point function for the $SU(4)$ theory is 
\begin{align}
\langle W_{1}\rangle^{SU(4)_{-5}}&=-q^{1/2}(-q;q)_{\infty}^2. 
\end{align}

We conjecture that for general $N$ we have
\begin{align}
\label{w1_suNkmN+1}
\langle W_{1}\rangle^{SU(N)_{-N-1}}(x;q)
&=\frac{1}{(q;q)_{\infty}}
\sum_{n\in \mathbb{Z}}
(-1)^{Nn+(N-1)}
q^{\frac{Nn^2}{2}+(N-1)n}
x^{Nn+(N-1)}. 
\end{align}

The expressions (\ref{h_suNkmN+1}) and (\ref{w1_suNkmN+1}) satisfy the $q$-difference equation
\begin{align}
\mathbb{II}_{\mathcal{N}}^{SU(N)_{-N-1}}(q^{(N-1)/N}x;q)
&=(-1)^{N-1}q^{-(N-1)/2}x^{-(N-1)}\langle W_{1}\rangle^{SU(N)_{-N-1}}(x;q). 
\end{align}
This should have an interpretation in terms of a dual description with a vortex line but we do not have a precise derivation of this equation -- we leave that for future work.

%%%%%%%%%%%%%%%%%%%%%%%%%%%%%%%%%%
\subsection{$k=-N-M$}
\label{sec_suN_kmNint}
%%%%%%%%%%%%%%%%%%%%%%%%%%%%%%%%%%
Now we can generalize the result for the $SU(N)_{k}$ pure Chern-Simons theory with $k\le -N-1$ being integer. 
This can be realized when $N_f=N_a=0$, $M\neq 0$ where the level is given by $k = -N - M$.   
The half-index is given by the matrix integral
\begin{align}
\label{h_suNmk_i}
\mathbb{II}_{\mathcal{N}}^{SU(N)_{-N - M}}(x_{\alpha};q)
&=
\frac{(q;q)_{\infty}^{N-1}}{N!}
\oint \left(
\prod_{i=1}^{N-1}
\frac{ds_i}{2\pi is_i}
\right)
\left(
\prod_{i\neq j}
(s_i s_j^{-1};q)_{\infty}
\right)
\prod_{i=1}^{N}
\prod_{\alpha=1}^{M}
(q^{\frac12}s_{i}^{\pm}x_{\alpha}^{\pm};q)_{\infty}, 
\end{align}
where $\prod_i s_i=1$ and $\prod_{\alpha}x_{\alpha}=1$. 

We conjecture that the half-index (\ref{h_suNmk_i}) is equal to 
\begin{align}
\label{h_wzwU(-k-N)_N}
\frac{1}{(q;q)_{\infty}^{M}}
\sum_{m_1,\cdots,m_{M} \in \mathbb{Z}}
\frac{(-1)^{N\sum_{\alpha=1}^{M}m_{\alpha}}
q^{\frac{N}{2}\sum_{\alpha=1}^{M}m_{\alpha}^2}
\prod_{\alpha=1}^{M}x_{\alpha}^{Nm_{\alpha}}
}
{
\prod_{\alpha<\beta}^{M}
(q^{1\pm m_{\alpha}\mp m_{\beta}} x_{\alpha}^{\pm}x_{\beta}^{\mp};q)_{\infty}
}. 
\end{align}
We have numerically checked the cases for $N=4$ and $N=5$ as well. 
The $q$-series (\ref{h_wzwU(-k-N)_N}) can be viewed as 
the vacuum character of the $U(M)_{N}$ WZW model 
so that the identity verifies the duality of boundary conditions proposed in \cite{Dimofte:2017tpi}
\begin{align}
&\textrm{$SU(N)_{-N - M}$ pure CS with Neumann b.c. $+$ $M$ fund. Fermis}
\nonumber\\
&\leftrightarrow 
\textrm{$U(M)_{N + M,N}$ pure CS with Dirichlet b.c.}
\end{align}

%%%%%%%%%%%%%%%%%%%%%%%%%%%%%%%%%%
\subsubsection{One-point function}
\label{sec_1pt_suN_kmNint}
%%%%%%%%%%%%%%%%%%%%%%%%%%%%%%%%%%
The one-point function of the Wilson line in the fundamental representation is given by
\begin{align}
\label{w1_suNmk_i}
&
\langle W_1\rangle^{SU(N)_{-N - M}}(x_{\alpha};q)
\nonumber\\
&=
\frac{(q;q)_{\infty}^{N-1}}{N!}
\oint \left(
\prod_{i=1}^{N-1}
\frac{ds_i}{2\pi is_i}
\right)
\left(
\prod_{i\neq j}
(s_i s_j^{-1};q)_{\infty}
\right)
\prod_{i=1}^{N}
\prod_{\alpha=1}^{M}
(q^{\frac12}s_{i}^{\pm}x_{\alpha}^{\pm};q)_{\infty} \sum_{i=1}^{N}s_i, 
\end{align}

We conjecture that 
the one-point function (\ref{w1_suNmk_i}) is equal to 
\begin{align}
&
\frac{(-1)^{N-1} q^{(N-1)/2}}
{(q;q)_{\infty}^{M}}
\sum_{m_1,\cdots,m_{M} \in \mathbb{Z}}
\frac{(-1)^{N\sum_{\alpha=1}^{M}m_{\alpha}}
q^{\frac{N}{2}\sum_{\alpha=1}^{M}m_{\alpha}^2}
\prod_{\alpha=1}^{M}x_{\alpha}^{Nm_{\alpha}}
}
{
\prod_{\alpha<\beta}^{M}
(q^{1\pm m_{\alpha}\mp m_{\beta}} x_{\alpha}^{\pm}x_{\beta}^{\mp};q)_{\infty}
}
\nonumber\\
&\times 
\sum_{
\begin{smallmatrix}
i_{j}\ge 0\\
i_1+i_2+\cdots i_{|k|-N}=N-1\\
\end{smallmatrix}
}
q^{i_1 m_1+i_2 m_2+\cdots +i_{M}m_{M}}
x_1^{i_1}x_2^{i_2}\cdots x_{M}^{i_{M}}. 
\end{align}
The one-point function (\ref{w1_suNmk_i}) can 
be obtained from the half-index (\ref{h_suNmk_i}) according to the $q$-difference equation
\begin{align}
&
\langle W_1\rangle^{SU(N)_{-N - M}}(x_1,\cdots, x_{M};q)
\nonumber\\
&=(-1)^{N-1}q^{(N-1)/2}
\sum_{
\begin{smallmatrix}
i_{j}\ge 0\\
i_1+i_2+\cdots i_{M}=N-1\\
\end{smallmatrix}
}
\left( \prod_{j = 1}^{M} x_j^{i_j} \right)
\mathbb{II}_{\mathcal{N}}^{SU(N)_{-N - M}}
\left(
q^{i_1/N}x_1,
q^{i_2/N}x_2, 
\cdots, 
q^{i_{M}/N}x_{M}
\right). 
\end{align}

%%%%%%%%%%%%%%%%%%%%%%%%%%%%%%%%%%
\subsection{$k=-N-M+1/2$}
\label{sec_suN_kmfrac}
%%%%%%%%%%%%%%%%%%%%%%%%%%%%%%%%%%
When we introduce a single fundamental chiral multiplet in the fundamental representation 
obeying the Neumann boundary condition as well as $M$ fundamental Fermi multiplets, 
we find the Chern-Simons theory with fractional level $k = -N - M + 1/2$. 
We have the half-index 
\begin{align}
\label{h_suN_kmfrac}
\mathbb{II}_{\mathcal{N}}^{SU(N)_k}(x_{\alpha};q)
&=
\frac{(q;q)_{\infty}^{N-1}}{N!}
\oint \left(
\prod_{i=1}^{N-1}
\frac{ds_i}{2\pi is_i}
\right)
\left(
\prod_{i\neq j}
(s_i s_j^{-1};q)_{\infty}
\right)
\prod_{i=1}^{N}
\prod_{\alpha=1}^{M}
\frac{(q^{\frac12}s_i^{\pm}x_{\alpha}^{\pm};q)_{\infty}}
{(q^{\frac12}s_i x_{M + 1};q)_{\infty}}, 
\end{align}
where $\prod_i s_i=1$ and $\prod_{\alpha}x_{\alpha}=1$. 

We conjecture that the integral expression (\ref{h_suN_kmfrac}) agrees with the series expresion
\begin{align}
\label{h_dualsuN_kmfrac}
&
\frac{1}{(q;q)_{\infty}^{M}}
\sum_{m_1,\cdots, m_{M}\in \mathbb{Z}}
\frac{
(-1)^{N\sum_{\alpha=1}^{M}m_{\alpha}}
q^{\frac{N}{2}\sum_{\alpha=1}^{M}m_{\alpha}^2}
\prod_{\alpha=1}^{M}x_{\alpha}^{Nm_{\alpha}}
}
{\prod_{\alpha<\beta}^{M} 
(q^{1\pm m_{\alpha}\mp m_{\beta}}x_{\alpha}^{\pm}x_{\beta}^{\mp};q)_{\infty}
}
\nonumber\\
&\times 
\prod_{\alpha=1}^{M}(q^{1-m_{\alpha}} x_{\alpha}^{-1}x_{M + 1};q)_{\infty}. 
\end{align}
This is consistent with the dual theory having Dirichlet boundary conditions for the $U(M)_{N 
+ M + 1/2, N + 1/2}$ vector multiplet and Dirichlet boundary conditions for the fundamental 3d chiral.
We have numerically confirmed the conjectural identity for $N=4$ and $N=5$ as well.

%%%%%%%%%%%%%%%%%%%%%%%%%%%%%%%%%%%
\subsection*{Acknowledgements}
%\textcolor{red}{The authors would like to thank XXX for useful discussions and comments.}
The work of T.O. was supported by the Startup Funding no.\ 4007012317 of the Southeast University. 
This work was supported in part by the STFC Consolidated grant ST/T000708/1.
%STFC Consolidated Grants ST/P000371/1 and ST/T000708/1.
%%%%%%

\appendix

%%%%%%%%%%%%%%%%%%%%%%%%%%%%%%%%%%
%%%%%%%%%%%%%%%%%%%%%%%%%%%%%%%%%%
\section{Special functions}
\label{sec_ramanujan}
%%%%%%%%%%%%%%%%%%%%%%%%%%%%%%%%%%
%%%%%%%%%%%%%%%%%%%%%%%%%%%%%%%%%%

%%%%%%%%%%%%%%%%%%%%%%%%%%%%%%%%%%
\subsection{Dedekind eta function}
\label{sec_eta}
%%%%%%%%%%%%%%%%%%%%%%%%%%%%%%%%%%
The Dedekind eta function is defined by
\begin{align}
\label{eta}
\eta(\tau)&:=q^{\frac{1}{24}}\prod_{n=1}^{\infty}(1-q^n)
\end{align}
where $q=e^{2\pi i\tau}$, $\tau\in \mathbb{H}$. 
It obeys 
\begin{align}
\eta(\tau+1)&=e^{\frac{\pi i}{12}}\eta(\tau),\\
\eta(-1/\tau)&=\sqrt{\tau/i}. 
\end{align}

%%%%%%%%%%%%%%%%%%%%%%%%%%%%%%%%%%
\subsection{Jacobi theta functions}
\label{sec_jacobi}
%%%%%%%%%%%%%%%%%%%%%%%%%%%%%%%%%%
The Jacobi theta functions are defined by
\begin{align}
\label{theta1}
\vartheta_1(z;\tau)
%&=
%-\vartheta_{1,1}(z;\tau)
%\nonumber\\
&=\sum_{n\in \mathbb{Z}}(-1)^{n-\frac12} q^{\frac12 (n+\frac12)^2}x^{n+\frac12}
=-\sum_{n\in \mathbb{Z}} e^{\pi i\tau (n+\frac12)^2+2\pi i(z+\frac12)(n+\frac12)}, \\
\label{theta2}
\vartheta_2 (z;\tau)
%&=
%\vartheta_{1,0}(z;\tau)
%\nonumber\\
&=\sum_{n\in \mathbb{Z}} q^{\frac12 (n+\frac12)^2}x^{n+\frac12}
=\sum_{n\in \mathbb{Z}} e^{\pi i\tau (n+\frac12)^2+2\pi iz(n+\frac12)}, \\
\label{theta3}
\vartheta_3(z;\tau)
%&=
%\vartheta_{0,0}(z;\tau)
%\nonumber\\
&=\sum_{n\in \mathbb{Z}} q^{\frac{n^2}{2}} x^n
=\sum_{n\in \mathbb{Z}} e^{\pi i\tau n^2+2\pi inz},\\
\label{theta4}
\vartheta_4(z;\tau)
%&=
%\vartheta_{0,1}(z;\tau)
%\nonumber\\
&=\sum_{n\in \mathbb{Z}} (-1)^n q^{\frac{n^2}{2}} x^n
=\sum_{n\in \mathbb{Z}} e^{\pi in^2+2\pi i(z+\frac12) n}
\end{align}
with $q=e^{2\pi i\tau}$, $x=e^{2\pi iz}$, $\tau\in \mathbb{H}$ and $z\in \mathbb{C}$. 

%%%%%%%%%%%%%%%%%%%%%%%%%%%%%%%%%%
\subsection{General Ramanujan's theta function}
\label{sec_gram}
%%%%%%%%%%%%%%%%%%%%%%%%%%%%%%%%%%
General Ramanujan's theta function is given by \cite{MR1117903}
\begin{align}
\label{r_theta}
f(a,b)
&=\sum_{m\in \mathbb{Z}}
a^{\frac{m(m+1)}{2}}b^{\frac{m(m-1)}{2}}
=(-a;ab)_{\infty}(-b;ab)_{\infty}(ab;ab)_{\infty}
\end{align}
where $|ab|<1$. 

For $|q|<1$ we introduce 
\begin{align}
\label{r_theta_phi}
\varphi(q)&:=f(q,q)=\sum_{n\in \mathbb{Z}}q^{n^2}=\frac{(-q;-q)_{\infty}}{(q;-q)_{\infty}}, \\
\label{r_theta_psi}
\psi(q)&:=f(q,q^3)=\sum_{n=0}^{\infty}q^{\frac{n(n+1)}{2}}=\frac{(q^2;q^2)_{\infty}}{(q;q^2)_{\infty}}, \\
\label{r_theta_f}
f(-q)&:=f(-q,-q^2)=\sum_{n\in \mathbb{Z}}(-1)^n q^{\frac{n(3n-1)}{2}}=(q;q)_{\infty}, \\
\label{r_theta_chi}
\chi(q)&:=(-q;q^2)_{\infty}. 
\end{align}

%%%%%%%%%%%%%%%%%%%%%%%%%%%%%%%%%%
\subsection{False general Ramanujan's theta function}
\label{sec_fgram}
%%%%%%%%%%%%%%%%%%%%%%%%%%%%%%%%%%
False general Ramanujan's theta function is given by \cite{rogers1917two}. 
\begin{align}
\label{r_theta_Psi}
\Psi(a,b):=\sum_{m=0}^{\infty}a^{\frac{m(m+1)}{2}}b^{\frac{m(m-1)}{2}}
-\sum_{m=1}^{\infty}a^{\frac{m(m-1)}{2}}b^{\frac{m(m+1)}{2}}. 
\end{align}
Note that 
unlike general Ramanujan's theta function, 
false general Ramanujan's theta function is not invariant under exchange of $a$ and $b$.

%%%%%%%%%%%%%%%%%%%%%%%%%%%%%%%%%%
%%%%%%%%%%%%%%%%%%%%%%%%%%%%%%%%%%
\bibliographystyle{utphys}
\bibliography{ref}

\end{document}